\documentclass[twocolumn]{aastex62}
\usepackage{amsmath}     
\usepackage{wasysym}           
\usepackage{graphicx}
\usepackage{amssymb}
\usepackage{epstopdf}
\usepackage{mathrsfs}
\usepackage{anyfontsize}
\usepackage{natbib}
\usepackage{color}
\usepackage{lipsum}
\usepackage{diagbox}

\shorttitle{Variability in Short Gamma-ray Bursts}
\shortauthors{E.~R.~Coughlin et.~al.}
\begin{document}
\title{Variability in Short Gamma-ray Bursts: Gravitationally Unstable Tidal Tails}
\author[0000-0003-3765-6401]{Eric R.~Coughlin}
\affiliation{Department of Astrophysical Sciences, Princeton University, Princeton, NJ 08544, USA}
\affiliation{Department of Physics, Syracuse University, Syracuse, NY 13244, USA}
\author[0000-0002-2137-4146]{C.~J.~Nixon}
\affiliation{School of Physics and Astronomy, University of Leicester, Leicester, LE1 7RH, UK}
\author{Jennifer Barnes}
\affiliation{Department of Physics and Columbia Astrophysics Laboratory, Columbia University, New York,
NY 10027, USA}
\author[0000-0002-4670-7509]{Brian D.~Metzger}
\affiliation{Department of Physics and Columbia Astrophysics Laboratory, Columbia University, New York,
NY 10027, USA}
\affiliation{Center for Computational Astrophysics, Flatiron Institute, New York, NY 10010, USA}
\author[0000-0003-4768-7586]{R.~Margutti}
\affiliation{Center for Interdisciplinary Exploration and Research in Astrophysics (CIERA) and Department of Physics and Astronomy, Northwestern University, Evanston, IL 60208}

\email{eric.r.coughlin@gmail.com}

\begin{abstract}
Short gamma-ray bursts are thought to result from the mergers of two neutron stars or a neutron star and stellar mass black hole. The final stages of the merger are generally accompanied by the production of one or more tidal ``tails'' of ejecta, which fall back onto the remnant-disc system at late times. Using the results of a linear stability analysis, we show that if the material comprising these tails is modeled as adiabatic and the effective adiabatic index satisfies $\gamma \ge 5/3$, then the tails are gravitationally unstable and collapse to form small-scale knots. We analytically estimate the properties of these knots, including their spacing along the tidal tail and the total number produced, and their effect on the mass return rate to the merger remnant. We perform hydrodynamical simulations of the disruption of a polytropic (with the polytropic and adiabatic indices $\gamma$ equal), $\gamma =2$ neutron star by a black hole, and find agreement between the predictions of the linear stability analysis and the distribution of knots that collapse out of the instability. The return of these knots to the black hole induces variability in the fallback rate, which can manifest as variability in the lightcurve of the GRB and -- depending on how rapidly the instability operates -- the prompt emission. The late-time variability induced by the return of these knots is also consistent with the extended emission observed in some GRBs. 
\end{abstract}


\section{Introduction}
Timing, energetics, and host galaxy (both the specific environments and lack of apparent proximity) constraints have suggested that short gamma-ray bursts (GRBs) originate from the merger of compact objects (e.g., \citealt{paczynski86, eichler89, burrows05, zhang07, berger14, fong15}). This theoretical notion was recently vindicated with the contemporaneous gravitational wave and gamma-ray observations of the event GW/GRB170817 \citep{abbott17}. Multiwavelength follow-up of this event also confirmed that such mergers produce a kilonova -- the radioactively powered emission from \emph{r}-process nucleosynthesis in the aftermath of the coalescence (e.g., \citealt{li98, metzger10,kasen17}). 

The emission from GRBs shows variability across a range of timescales (e.g., \citealt{margutti11, dichiara13, berger14, swenson14, mu18}). The origin of this variability could be related to an intrinsic restructuring of the accretion disc surrounding the post-merger object \citep{perna06,dall17}. Another possibility, however, is related to the fallback of material onto the post-merger system that occurs as the tidal tails of debris -- ejected during the final stages of the inspiral -- rain back onto the disc. In particular, while it is predicted that the overall scaling of this fallback should trace a smooth, $\propto t^{-5/3}$ decline in time (e.g., \citealt{chevalier89}; see also the fallback rate in Figure \ref{fig:fallback} below), small scale structure in the debris that feeds the accretion flow could correspondingly lead to changes in the luminosity of the system. 

Such structure could be supplied by the intrinsic nature of the object(s) destroyed in the merger; for example, convection in the interior of a neutron star (e.g., \citealt{thompson93}), which may be enhanced during the inspiral, will naturally provide local density fluctuations in the tails of the debris. It is also conceivable that the dynamical state of the ejected material is susceptible to a larger-scale instability, leading to the formation of knots within the tails that then return at discrete times. \citet{rosswog07} speculated about this latter possibility, and \citet{lee07} found from their numerical simulations that such an instability could indeed occur (see also \citealt{colpi94, rasio94, lee00}). However, \citet{lee07} only found that the tails were unstable if the polytropic index of the gas comprising the tails satisfied $\gamma \gtrsim 3$, whereas more realistic equations of state likely yield $\gamma \sim 2 - 3$ (e.g., \citealt{rasio94, lattimer01}). 

Here we use a combination of analytical arguments and numerical experiments to demonstrate that the tails generated from compact object mergers are gravitationally unstable -- and collapse to form small scale knots that lead to variability in the fallback rate -- provided that the adiabatic index of the gas comprising the tails satisfies $\gamma \ge 5/3$. In Section \ref{sec:analytics} we provide analytic estimates of the linear growth rate of the instability and the properties of the knots that condense out of the tail, and in Section \ref{sec:numerics} we compare these predictions to numerical hydrodynamical simulations. We discuss the observational implications of these findings in Section \ref{sec:observational}. We summarize and conclude in Section \ref{sec:summary}.

\section{Stability analysis and analytic estimates}
\label{sec:analytics}
Polytropic, hydrostatic cylinders are gravitationally unstable to perturbations along the axis of the cylinder below a critical wavenumber, $k_{\rm crit}$, where $k_{\rm crit} \sim few$ is measured in units of the radius of the cylinder \citep{coughlin20}. By polytropic we mean that the adiabatic index of the gas, which controls how vigorously pressure perturbations respond to density perturbations, and the polytropic index, being the exponent $\gamma$ that appears in the relation $p \propto \rho^{\gamma}$, where $p$ is the pressure and $\rho$ is the density that appear in the equation of hydrostatic balance, are equal. All perturbations with wavenumbers below $k_{\rm crit}$ are unstable and grow as $e^{\sigma \tau}$, where $\sigma$ is the growth rate that depends on the wavenumber and $\tau$ is time in units of the sound crossing time over the radius of the cylinder, and cause runaway collapse along the cylinder axis. However, there is a second wavenumber $k_{\rm max} \sim 1$ ($< k_{\rm crit}$) at which the growth rate of the instability is maximized at a value of $\sigma_{\rm max} \sim 1$ (Figure 4 of \citealt{coughlin20}). If a hydrostatic cylinder is subjected to a random initial perturbation, such that the Fourier coefficients have comparable power over the range of unstable wavenumbers, then this wavelength will grow fastest and will characterize the mass and separation scales of the objects that condense out of the instability.

The formation of a cylindrical filament is a natural consequence of the tidal stretching of the debris that is flung out during the final stages of the merger of two compact objects (e.g., Figure \ref{fig:disruption}).  The difference between such a tidal tail and a hydrostatic cylinder is that the former possesses a non-negligible amount of shear in the velocity along the axis of the cylinder, which is established by the tidal field of the disrupting object and inhibits the growth of the instability. In the next subsection we analyze the  general behavior of the instability in the presence of shear. In Section \ref{sec:fragmentation} we limit the analysis to a polytropic index of $\gamma =2$ and we present the observational implications of the instability. 

\subsection{General stability analysis}
The tail expands predominantly in one direction as a consequence of the tidal stretching, which we define as the $z$-direction and delimits the axis of the cylinder, and we define $s$ as the cylindrical-radial direction and $\varphi$ as the azimuthal angle around the axis of the cylinder. Along $z$ there is a location $Z(t)$ where the material is marginally bound to the remnant, and the fluid element at this location therefore obeys 

\begin{equation}
\frac{\partial Z}{\partial t} \equiv V_{\rm Z} = \sqrt{\frac{2GM_{\bullet}}{Z}}, \label{dZdt}
\end{equation}
where $M_{\bullet}$ is the mass of the remnant. Near this marginally bound position we can Taylor expand the fluid variables in the quantity

\begin{equation}
\frac{z-Z(t)}{Z(t)} \equiv \frac{\Delta z}{Z(t)},
\end{equation}
and the leading-order (in $\Delta z/Z$) solution to the fluid equations for the $z$-component of the velocity is \citep{coughlin16}

\begin{equation}
v_{\rm z} = V_{\rm Z}\left(1+2\frac{\Delta z}{Z}\right). \label{vzapp}
\end{equation}
If we now assume that the gas is isentropic and we let the adiabatic index satisfy $\gamma \ge 5/3$, then we can show that the leading-order solutions to the fluid equations for the other fluid variables are

\begin{equation}
\rho = \frac{\Lambda(Z)}{4\pi H(Z)^2}g(\xi), \,\,\, p = \frac{G\Lambda^2}{4\pi H^2}h(\xi), \,\,\,v_{\rm s} = V_{\rm H}\xi, \label{sols}
\end{equation}
where $\rho$ is the gas density, $p$ is the pressure, $v_{\rm s}$ is the cylindrical-radial velocity, $H(Z)$ is the radius of the cylinder where the density equals zero, $V_{\rm H} = \partial H/\partial t$ is the surface velocity that satisfies

\begin{equation}
V_{\rm H} = \frac{2-\gamma}{\gamma-1}V_{\rm Z}\frac{H}{Z}
\end{equation}
and hence $H \propto Z^{(2-\gamma)/(\gamma-1)}$, $\xi = s/H$, and

\begin{equation}
\Lambda \propto \int_0^{H}\rho\,s\,ds \propto Z^{-2}
\end{equation}
is the line mass of the cylinder. With these definitions and the isentropic relation between the pressure and the density, so that $h = K g^{\gamma}$ with $K$ the entropy, we can show that the cylindrical component of the momentum equation and the Poisson equation can be combined to give

\begin{equation}
\frac{K\gamma}{\gamma-1}\frac{\partial}{\partial \xi}\left[\left(\frac{1}{\xi}\frac{\partial \lambda}{\partial \xi}\right)^{\gamma-1}\right] = -\frac{\lambda}{\xi}, \label{hse}
\end{equation}
where $\lambda = \int_0^{\xi}g(\xi)\,\xi\,d\xi$ is the dimensionless line mass and $\lambda(1) = 1$. Equation \eqref{hse} is the Lane-Emden equation for the line mass of the cylinder as a function of cylindrical radius, and can be integrated and solved for $K$ as described in \citet{coughlin20} (see their Figure 1). 

The solutions given by Equation \eqref{sols} are quasi-hydrostatic in that the fluid velocity is non-zero in the cylindrical-radial direction but the sound speed $c_{\rm s}$ declines less rapidly than $V_{\rm H}$; specifically, we have

\begin{equation}
c_{\rm s} \simeq \sqrt{G\Lambda} \propto Z^{-1}, \label{cs}
\end{equation}
while the velocity at the surface of the tail obeys

\begin{equation}
V_{\rm H} \propto Z^{\frac{2-\gamma}{\gamma-1}-\frac{3}{2}}.
\end{equation}
Thus, as time advances and the tail continues to stretch, the propagation speed of perturbations within the tail is given by the sound speed and the relevant, dimensionless timescale that characterizes the evolution of any perturbation is

\begin{equation}
d\tau \simeq \frac{c_{\rm s}}{H}dt \equiv \frac{\sqrt{G\Lambda}}{H}dt. \label{dtau}
\end{equation}
With this dimensionless timescale, we can now introduce perturbations to the density, pressure, and gravitational potential that are functions of $\xi = s/H$, $\eta = \Delta z/H$, $\varphi$, and $\tau$ and derive the linearized perturbation equations from the fluid equations. When $\gamma > 5/3$, the resulting set of equations is identical to that derived in \citet{coughlin20} because of the quasi-hydrostatic nature of the unperturbed solutions, and there is correspondingly a maximally growing and unstable mode that will characterize the properties of the knots that collapse out of the ejected tail(s). When $\gamma = 5/3$, the analysis and the linearized perturbation equations are more complicated owing to the identical scaling between the sound speed and the expansion rate of the surface, and there is an additional parameter -- being the ratio of the sound speed to the expansion speed, which can be rewritten as the ratio of the tail density to the black hole tidal density\footnote{By the tidal density of the black hole we mean $ M_{\bullet}/r^3$.} -- that enters into the equations. When the tail density is much larger than the black hole density, then the equations reduce to those derived in \citet{coughlin20} and the instability grows as a power-law in time with the power-law index appropriate to Figure 4 in that paper. However, when the ejecta and black hole densities are comparable, the unstable eigenvalue must be derived as a function of the ratio of those densities. 

Here we focus on the case when $\gamma > 5/3$, as this has been the focus of past investigations of neutron star mergers, is likely appropriate for the very stiff equations of state that characterize nuclear densities (valid for the initial expansion of the tidal tail; see Section \ref{sec:thermal} below), and correspondingly the results directly carry over from \citet{coughlin20} with an appropriate redefinition of the dimensionless variables.  Specifically, Equation \eqref{dtau} combined with the scalings for the line mass of the stream and the stream width gives

\begin{equation}
\begin{split}
\tau = \frac{\sqrt{G\Lambda_0}}{H_0}\frac{Z_0^{3/2}}{\sqrt{2GM_{\bullet}}}&\frac{1}{\frac{3}{2}-\frac{1}{\gamma-1}}  \\ 
&\times\left(\left(\frac{Z}{Z_0}\right)^{\frac{3}{2}-\frac{1}{\gamma-1}}-1\right), \label{tauofZ}
\end{split}
\end{equation}
where $\Lambda_0$, $H_0$, and $Z_0$ are the initial line mass, stream radius, and location of the marginally-bound orbit, respectively. If we further integrate Equation \eqref{dZdt} to give

\begin{equation}
\frac{Z}{Z_0} = \left(1+\frac{3}{2}\frac{\sqrt{2GM_{\bullet}}}{Z_0^{3/2}}t\right)^{2/3}, \label{Zoft}
\end{equation}
and use the fact that the object(s) that gave rise to the tidally disrupted tail(s) had a pericenter distance comparable to the tidal radius, so that $Z_0^{3/2}/\sqrt{2GM_{\bullet}} \simeq H_0/\sqrt{G\Lambda_0}$, then for $t \gtrsim t_0$ Equation \eqref{tauofZ} becomes

\begin{equation}
\tau \simeq \left(\frac{t}{t_0}\right)^{1-\frac{2}{3}\frac{1}{\gamma-1}},
\end{equation}
where 

\begin{equation}
    t_0 = \frac{2}{3}\frac{R_\star^{3/2}}{\sqrt{2GM_{\star}}}
\end{equation}
is roughly the sound crossing time over the original star of radius $R_\star$ and mass $M_\star$.

The existence of the unstable mode implies that perturbations to the density at the maximally growing mode increase with time as $\simeq e^{\sigma_{\rm max}\tau}$, which is exponential to a fractional power of time, and the fractional power depends on the adiabatic index but approaches 1 as $\gamma$ becomes large. If we denote the dimensionless amplitude of the seed perturbation at the maximally growing mode by $\delta\rho_0$, then the time dependent evolution of the perturbation to the density at the wavelength $k_{\rm max}$ is 

\begin{equation}
    \frac{\delta\rho}{\rho_0} \simeq \delta\rho_0 e^{\sigma_{\rm max}\tau},
\end{equation}
where $\rho_0$ is the background density (which is declining owing to the stretching of the tail). The nonlinear phase of the instability is reached when $\delta \rho \simeq \rho_0$, which is correspondingly near the time at which we expect the tail to fragment. 

We note that the analysis here assumed that the axis of the cylinder was confined to a single direction (which we defined as the $z$-direction), whereas in reality the tidal tail ejected from a merger will possess some curvature that results from the non-zero angular momentum of the debris (see Figure \ref{fig:disruption}). This additional aspect of the problem can be trivially incorporated into the perturbation analysis, as such curvature amounts to a $z$-dependent displacement of the axis of the cylinder, which corresponds to an $m = 1$ perturbation (where perturbations in the $\phi$ direction around the axis of the cylinder vary as $\propto e^{im\phi}$ with $m$ an integer; an $m =1$ perturbation displaces the axis of the cylinder, analogous to the way in which an $\ell = 1$ perturbation of a spherical configuration of gas displaces the center of mass of the sphere). However, it can be shown that such perturbations are stable \citep{breysse14,coughlin20}, and hence the curvature of the tidal tail has no impact on the linear growth of the instability.

\subsection{Fragmentation timescales and observational implications for short GRBs}
\label{sec:fragmentation}

Taking an adiabatic index of $\gamma = 2$ for concreteness, the time at which this nonlinear phase is reached, which we denote $t_{\rm frag}$, is (inverting the above expressions for the density and $\tau$)

\begin{equation}
    t_{\rm frag} \simeq t_0\frac{\ln^3\left(\frac{ 1}{\delta\rho_0}\right)}{\sigma_{\rm max}^3}. \label{tfrag0}
\end{equation}
As discussed in more detail below (see Figure \ref{fig:rhoi_of_tau} and the discussion thereof), the initial amplitude of the perturbation at the maximally growing mode will be small if the density profile of the neutron star is as smooth as a $\gamma = 2$ polytrope, and nonlinear couplings between smaller-$k$ perturbations are likely responsible for the emergence of the most unstable mode. If we therefore take $\delta\rho_0 \simeq 10^{-3}$, $\sigma_{\rm max} = 0.57$ (Table 2 of \citealt{coughlin20}) and -- with $R_\star = 11$ km and $M_{\star} = 1.5M_{\odot}$ -- $t_0  \simeq 0.04$ ms, then this timescale is

\begin{equation}
    t_{\rm frag} \simeq 70 \textrm{ ms}. \label{tfrag1}
\end{equation}
At approximately this time we expect the stream to fragment into $N$ knots separated by a spacing of $z_{\rm sep} \simeq 2\pi R_\star/k_{\rm max} \simeq 30$ km along the tail and near the marginally bound radius, where $k_{\rm max} \simeq 0.96$ is the wavenumber of the maximally growing mode in units of the cylindrical radius of the tail (see Table 2 of \citealt{coughlin20}). We can estimate the number $N$ by noting that at this time the length of the stream $L$ has expanded by an amount $L \propto Z^2$, and hence, using Equation \eqref{Zoft}, 

\begin{equation}
   N_{\rm upper} \simeq \frac{L}{z_{\rm sep}} \simeq \frac{k_{\rm max}}{2\pi}\frac{\ln^4\left(\frac{1}{\delta\rho_0}\right)}{\sigma_{\rm max}^4} \simeq 3\times 10^3. \label{Nupper}
\end{equation}
This is, however, an overestimate of the true number, owing to the fact that a substantial fraction of the material will have already returned to the black hole by this time. A simple approximation of the return time of the most bound debris can be calculated from the tidal and frozen-in approximations \citep{lacy82, rees88, lodato09, stone13}, which gives 

\begin{equation}
    T_{\rm fb} \simeq \left(\frac{R_\star}{2}\right)^{3/2}\frac{2\pi M_{\bullet}}{M_{\star}\sqrt{GM_{\bullet}}} \simeq 0.5\textrm{ ms},
\end{equation} 
where we set $M_{\bullet} = 5M_{\odot}$ for the black hole mass, though the comparable size of the neutron star and the black hole implies that material is much deeper in the potential well initially and is promptly accreted (see the hydrodynamical simulations below). Thus, assuming that all of the stream is available to fragment results in a large overestimate of the number of knots for the types of encounters considered here. We can determine a lower bound on the number of knots formed out of the instability by taking the length of the stream to be of the order the marginally bound radius, which gives

\begin{equation}
    N_{\rm lower} \simeq \frac{Z}{z_{\rm sep}} \simeq \frac{k_{\rm max}}{2\pi}\frac{\ln^2\left(\frac{1}{\delta\rho_0}\right)}{\sigma_{\rm max}^2} \simeq 20. \label{Nlower}
\end{equation}
In general we expect the number of fragments to be between these upper and lower limits, and the exact number will depend on the disruption dynamics and the mass ratios involved (though the lower limit is likely more realistic for these systems where the return time of the most bound debris is much shorter than the fragmentation time). Depending on the amount of mass contained in the ejecta, we therefore expect the masses of the fragments to be on the order of $10^{-3} - 10^{-4} M_{\star}$, and thus $\sim 10^{-3}\,M_{\odot}$ .

\begin{figure}
    \centering
    \includegraphics[width=0.49\textwidth]{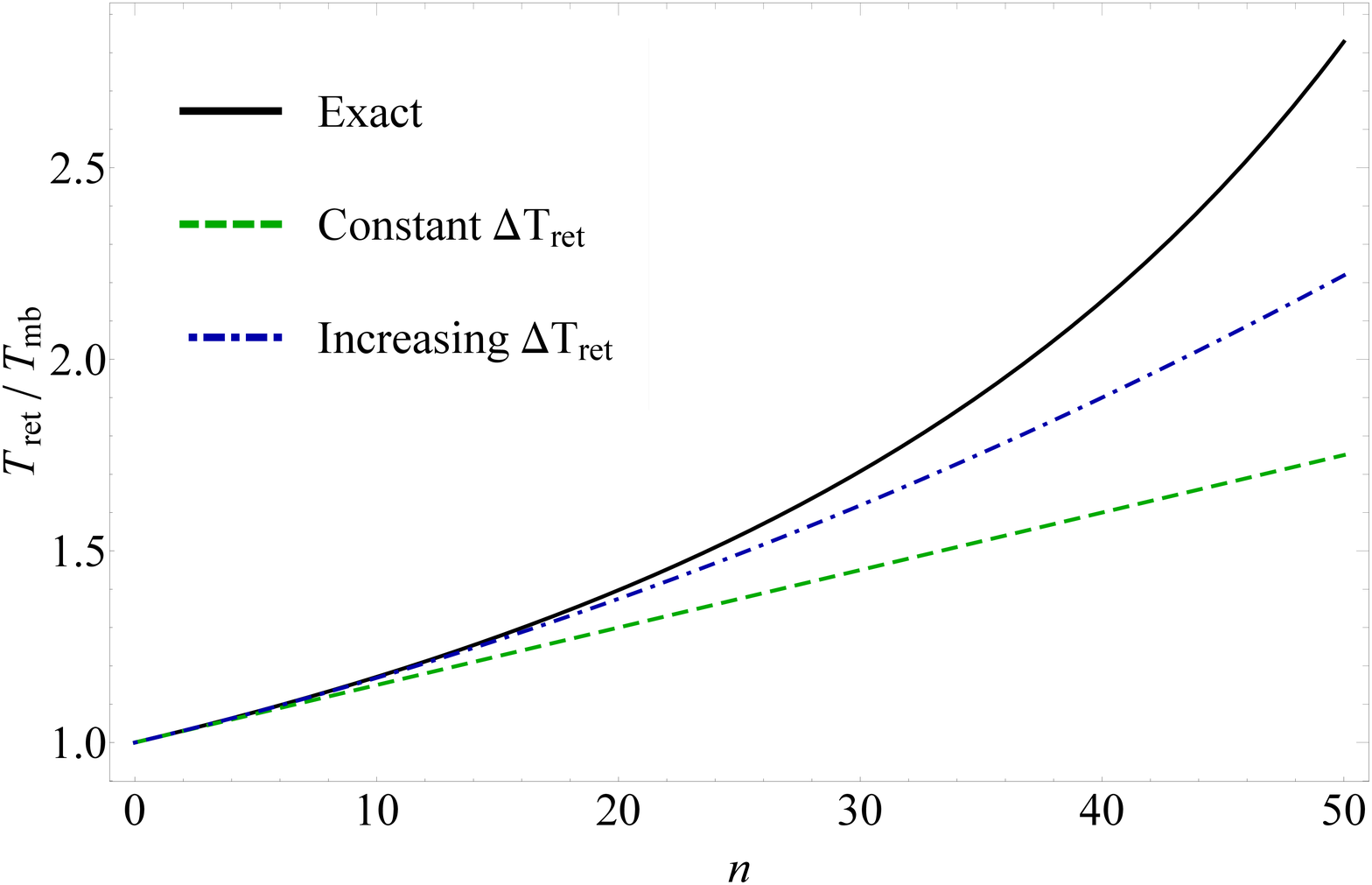}
    \caption{An example of the return time of the $n^{\rm th}$ knot $T_{\rm ret}$, normalized by the return time of the most-bound knot $T_{\rm mb}$, when the knots are separated along the stream by $z_{\rm sep} = 0.01\Delta Z_0$, where $\Delta Z_0$ is the distance between the most bound knot and the marginally bound radius at the time that fragmentation occurs. The black curve gives the exact solution (Equation \ref{Tret}), the green, dashed curve shows the leading-order solution for which the return time of successive knots is constant, and the blue, dot-dashed curve contains the higher-order correction that accounts for the increase in the return time between knots as the number of knots $n$ increases.} 
    \label{fig:Tret}
\end{figure}

After the fragments form we can approximate their dynamical evolution by assuming that they evolve as point masses purely in the gravitational field of the black hole. Using Equation \eqref{vzapp}, the Keplerian energy $\epsilon$ of the knots at the time they form is given by

\begin{equation}
    \epsilon = \frac{1}{2}v^2-\frac{GM_{\bullet}}{z} \simeq \frac{5GM_{\bullet}}{Z}\frac{\Delta z}{Z},
\end{equation}
where $\Delta z$ is the distance of a given fragment from the marginally bound position $Z$ at that time. With the energy-period relation of a Keplerian orbit, we find that the return time of the knots within the tail is

\begin{equation}
    T_{\rm ret} = 2\pi\left(\frac{Z}{10}\right)^{3/2}\frac{1}{\sqrt{GM_{\bullet}}}\left(-\frac{Z}{\Delta z}\right)^{3/2}.
\end{equation}
If we now define the distance of the most bound knot from the marginally bound radius by $-\Delta Z_{0}$, so that the distance of the $n^{\rm th}$ knot along the tail from the most bound knot is $-\Delta Z_0+n z_{\rm sep}$, then the $n^{\rm th}$ knot returns at a time

\begin{equation}
    T_{\rm ret}(n) = T_{\rm mb}\left(1-n\frac{z_{\rm sep}}{\Delta Z_0}\right)^{-3/2}, \label{Tret}
\end{equation}
where

\begin{equation}
    T_{\rm mb} = 2\pi\left(\frac{Z}{10}\right)^{3/2}\frac{1}{\sqrt{GM_{\bullet}}}\left(\frac{Z}{\Delta Z_0}\right)^{3/2}
\end{equation}
is the return time of the most bound knot, i.e., $T_{\rm mb} = T_{\rm ret}(n = 0)$. Adopting the limit of $z_{\rm sep} \ll Z$, which is valid when the number of knots formed is large, the temporal spacing between the return of successive knots $n+1$ and $n$ is

\begin{equation}
    \Delta T_{\rm ret} \simeq \frac{3}{2}T_{\rm mb}\frac{z_{\rm sep}}{\Delta Z_0}\left\{1+\frac{5}{2}n\frac{z_{\rm sep}}{\Delta Z_0}+\mathcal{O}\left(\frac{z_{\rm sep}^2}{\Delta Z_0^2}\right)\right\}. \label{dTret}
\end{equation}
This expression demonstrates that, while $n z_{\rm sep}/\Delta Z_0 \lesssim 1$, the spacing between the return of successive knots is roughly constant; using the numbers derived above and setting $\Delta Z_0 \simeq 0.1 Z$ at $t_{\rm frag}$, $T_{\rm mb} \simeq 1$ s, and this temporal spacing is $\sim 100$ ms. However, as the number of knots to have fallen back increases, the temporal spacing between successive knots increases owing to the reduced binding energy of knots initially at a larger distance, and the fractional increase is $\sim 5 z_{\rm sep}/(2\Delta Z_0)$. Thus, while the temporal spacing between the first and second knot is $\sim 100$ ms, the spacing between the 9th and the 10th is $\sim 240$ ms, the 99th and 100th is $\sim 2.6$ s, etc. As an example, Figure \ref{fig:Tret} illustrates the ratio of the return time $T_{\rm ret}$ to the return time of the most-bound knot $T_{\rm mb}$ for $z_{\rm sep}/\Delta Z_0 = 0.01$ as a function of the number of knots to have returned to pericenter $n$. The red curve shows the exact dependence, Equation \eqref{Tret}, the green, dashed curve gives the leading-order solution with a constant return time between knots, and the blue, dot-dashed curve contains the higher-order correction that accounts for the secular increase in return time between knots.

These estimates predict that, if the variability in the lightcurve of a short GRB traces the fallback of material to the compact object and the tail is gravitationally unstable to this mechanism, then the spacing between outbursts should be roughly evenly spaced at early times, but should lengthen approximately linearly with the number of flares observed. However, these estimates assume that the tail fragments precisely at the maximally growing mode and generates a specific separation between the resulting knots, while in reality there will be a range of separations owing to the fact that the tail is unstable to a continuum of wavelengths below a critical one. Nonetheless, it is likely that there will be enhanced power in the lightcurve at a frequency corresponding to the inverse of $\Delta T_{\rm ret}$ given in Equation \eqref{dTret}.

In the next section we compare these analytic predictions to hydrodynamical simulations.

\section{Hydrodynamical Simulations}
\label{sec:numerics}
\subsection{Initial setup}
\label{sec:initial}
We use the smoothed particle hydrodynamics (SPH) code {\sc phantom} \citep{price18} to simulate the disruption of a $2.0 M_{\odot}$ neutron star (NS) by a $5 M_{\odot}$ black hole (BH). We use a Paczy\'nsky-Wiita (PW) potential \citep{paczynsky80} to model the gravitational field of the BH, and we set its accretion radius -- inside of which particles are ``accreted'' and removed from the simulation -- to its Schwarzschild radius. The NS is modeled as a pure polytrope with a radius of 11 km, and we focus primarily on the case where the polytropic index (equal to the adiabatic index) is $\gamma = 2$. We maintain a polytropic equation of state for the entire duration of the simulation, i.e. the pressure is related to the density by $P=K\rho^\gamma$ where both $K$ and $\gamma$ are global fixed constants\footnote{Note that, since the code is Lagrangian, this is identical to solving the entropy equation in the absence of shocks.}, which implies that any heat generated from viscosity or shocks is lost from the system. Our fiducial resolution is $N_{\rm p} \simeq 1.4\times 10^{7}$ SPH particles. All other aspects of the modeling as related to the code (e.g., the implementation of self-gravity) are identical to those described in \citet{coughlin15}. In Appendix \ref{sec:caveats} we discuss the caveats of our numerical approach and speculate as to the effects of relaxing some of our assumptions (specifically, the disc physics in Section \ref{sec:disc}, the inclusion of general relativity in Section \ref{sec:gr}, the microphysics and thermal physics of the tidal tails in Section \ref{sec:thermal}, the numerical resolution in Section \ref{sec:resolution}, and the variation of the simulation parameters in Section \ref{sec:orbital}).

We do not model the inspiral of the binary system from well outside the tidal disruption radius of the NS, which would, at the very least, require the inclusion of post-Newtonian terms and an accurate modeling of the tidal dissipation in the star. Instead, we initialize the SPH particles with the velocity of the center of mass of the star, which itself is calculated to reproduce a parabolic orbit with a pericenter distance equal to the tidal disruption radius using the PW potential. Due to the nature of the encounter (near equal mass, close approach) the usual approach to estimating the tidal radius -- at which the star is completely disrupted -- is inaccurate. Instead we experimented with the pericenter distance for the NS orbit and found that a pericenter of $3GM_{\bullet}/c^2$ (where $M_{\bullet}$ is the mass of the black hole) is the largest pericentre that yields a fully disrupted NS and thus an appropriate debris stream. Such a ``plunge'' into the tidal radius may happen naturally on a single passage in a very dense stellar environment such as a globular cluster (e.g., \citealt{rosswog09}), or may arise from the rapid (i.e., sub-orbital-time) inspiral that accompanies the final stages of the tidal and gravitational-wave induced coalescence \citep{lai94, rasio94}. Our motivation for this approach parallels the focus of this paper, which is to assess the gravitational stability of the tidal tails formed as a byproduct of the inspiral, and not to understand the precise mechanism of the production itself. A number of other authors (e.g., \citealt{lee07, bauswein13, kyutoku13, brege18, foucart19}) have demonstrated that tidal tails are a fairly generic consequence of inspiral events with more realistic initial conditions (see also Section \ref{sec:orbital} below).

\subsection{Results}

\begin{figure*}[htbp] 
   \centering
   \includegraphics[width=0.495\textwidth]{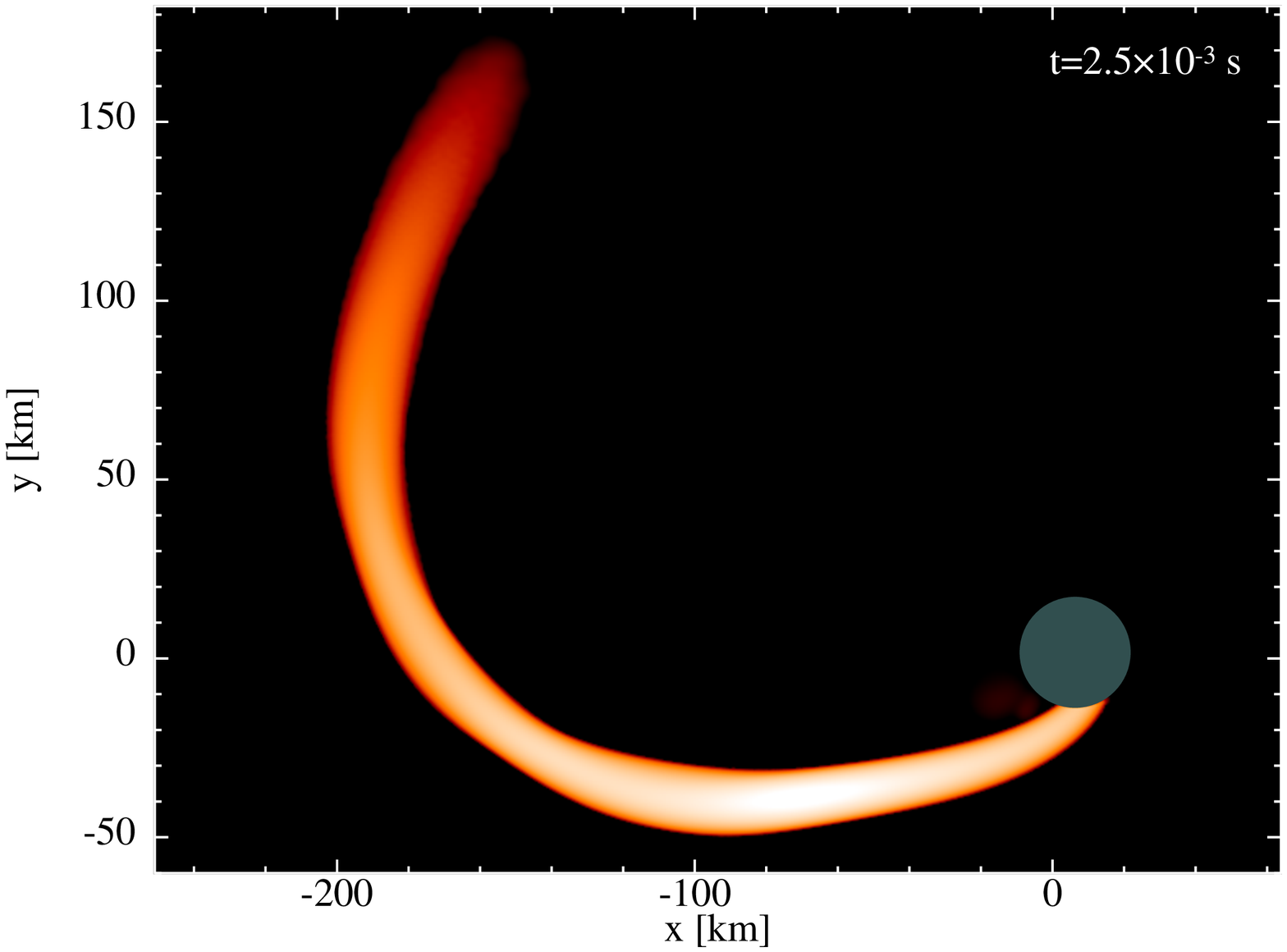} 
   \includegraphics[width=0.495\textwidth]{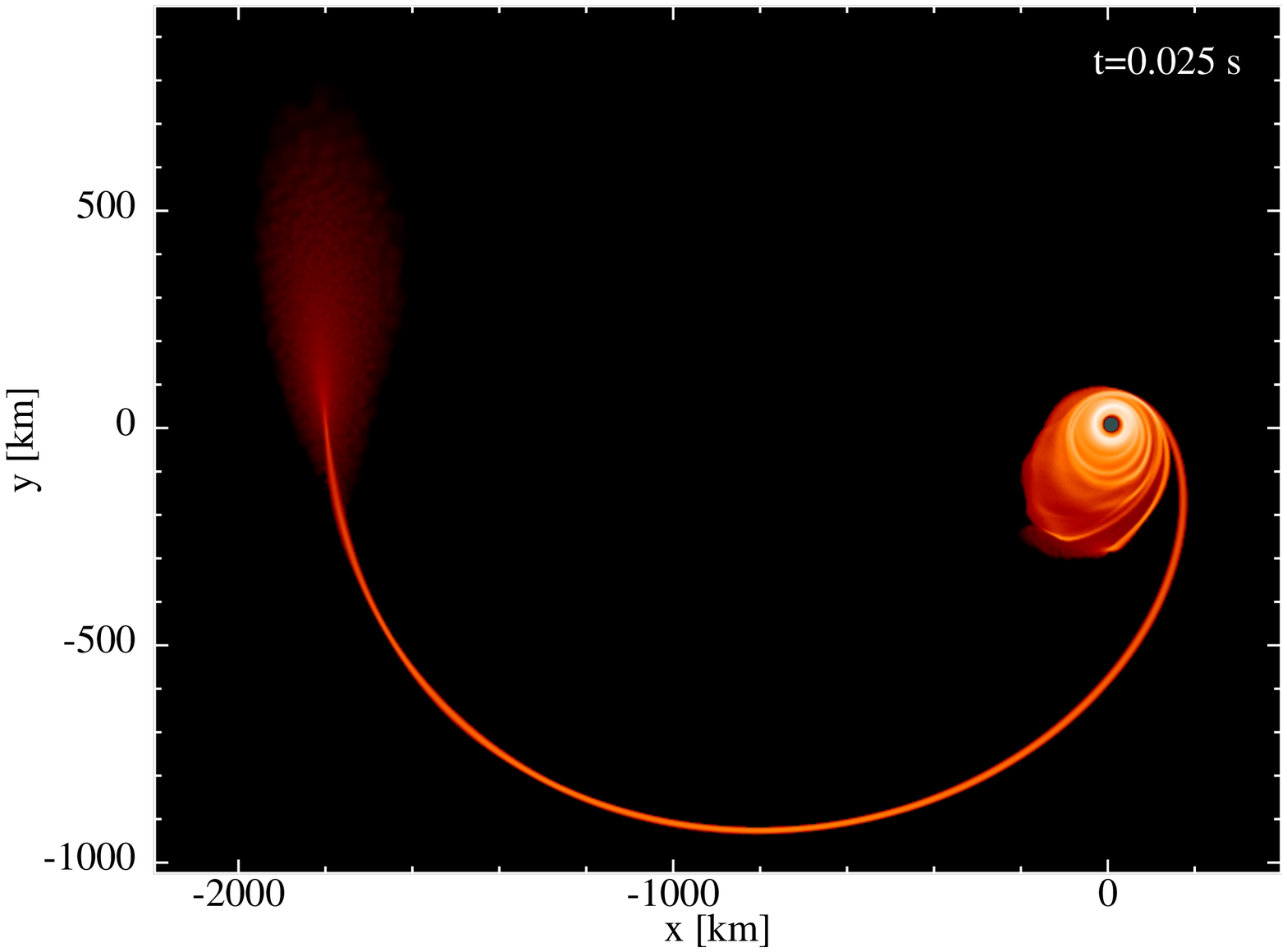}\vspace{0.1in} 
   \includegraphics[width=1.0\textwidth]{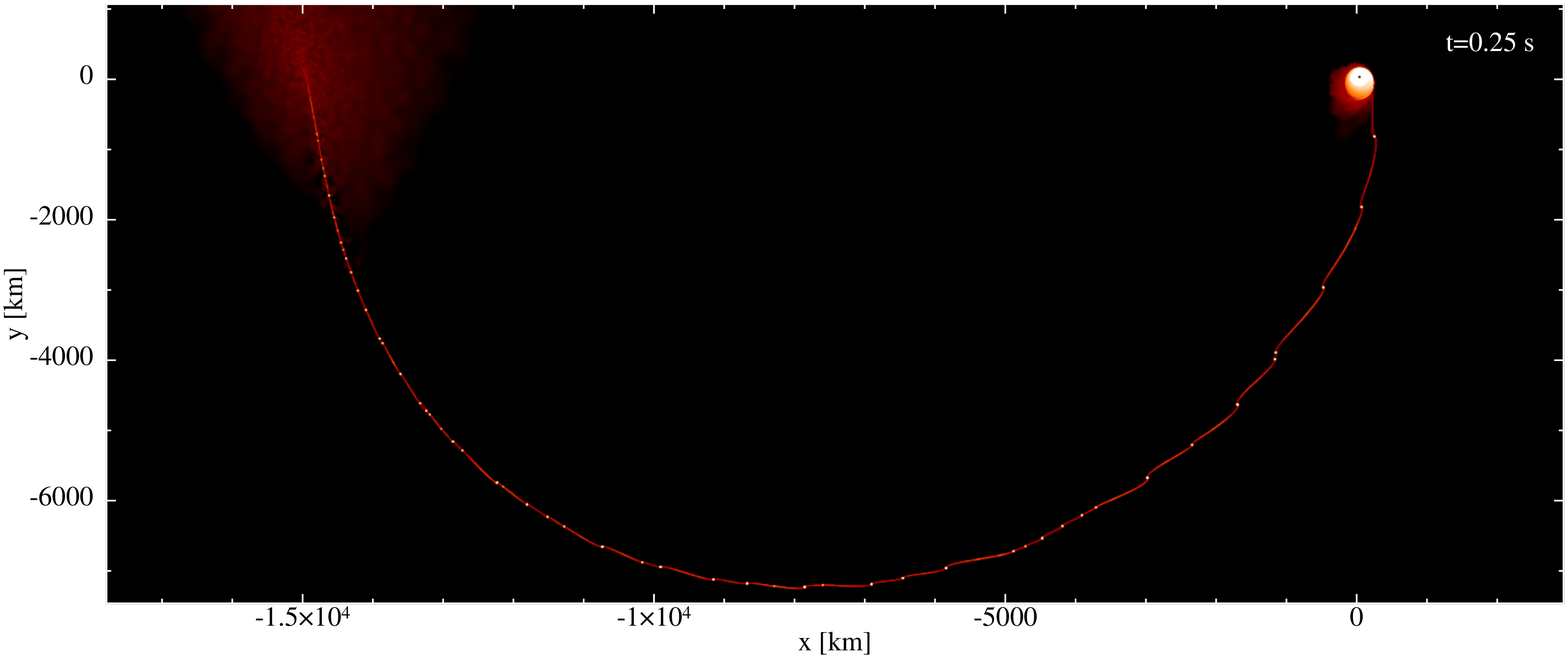} 
   \caption{The morphology of the disrupted debris following the tidal destruction of a neutron star by a black hole. Here the mass of the neutron star is $2M_{\odot}$ with a radius of 11 km, and the black hole (shown by the gray circle) has a mass of $5M_{\odot}$ and an accretion radius -- inside of which SPH particles are ``accreted'' and removed from the simulation -- equal to its Schwarzschild radius (see Section \ref{sec:initial} for more details of this specific simulation). Each panel corresponds to a different time post-disruption, as shown in the top-right corner of each panel. Colors indicate the column density projected onto the plane of the orbit of the original star, with brighter (darker) colors indicating regions of enhanced (reduced) density. The tidal potential of the black hole causes a large fraction ($\gtrsim 90\%$) of the neutron star to immediately accrete onto the black hole, which forms a prompt accretion flow, while $\sim 10\%$ of the mass is launched out from the system in the form of a tidal tail. At late times ($\sim few\times 0.1$ s) this ejected tail fragments under its own self-gravity, forming knots that return to and impinge upon the black hole-disc system, as can be seen in the bottom panel.}
   \label{fig:disruption}
\end{figure*}

Figure \ref{fig:disruption} shows the morphology of the disrupted debris following the tidal interaction between the neutron star and the black hole; here the column density of the material projected onto the plane of the binary is plotted, with brighter regions indicating denser material. The top-left, top-right, and bottom panels show the distribution of the material immediately post-pericenter, after the disc has formed but the tidal tail remains smooth, and after the tail has fragmented, respectively.  This figure demonstrates that, as a consequence of the energy spread imparted by the tidal potential of the black hole, most of the neutron star material ($\gtrsim 90\%$ of the initial mass of the star) is tightly bound to the black hole and is either swallowed directly or promptly forms an accretion disc. However, $\sim 10\%$ of the initial neutron star mass is ejected in the form of a tidal tail that, after many dynamical times at the tidal radius (note that $GM_{\bullet}/c^3 \simeq 2.5\times 10^{-5}$ s here), fragments and falls back onto the black hole-disc system.

\begin{figure}[htbp] 
   \centering
   \includegraphics[width=0.47\textwidth]{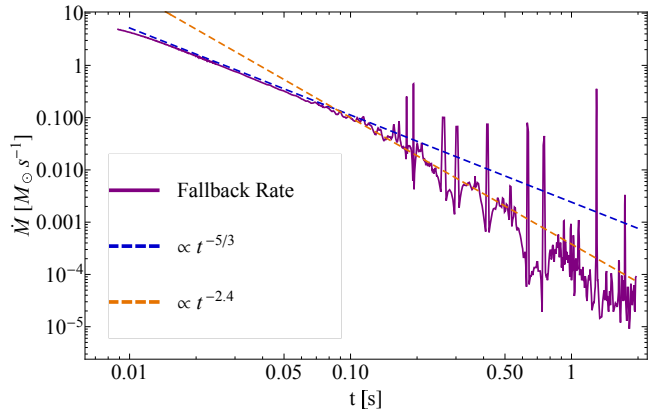} 
   \caption{The fallback rate onto the black hole from the returning tail in units of Solar masses per second as a function of time in seconds. After a time of $\sim 0.1$ s, the fallback rate starts to exhibit variability from the return of the knots that have gravitationally condensed out of the stream. The blue, dashed curve shows the $\propto t^{-5/3}$ scaling predicted for material returning from the marginally-bound radius within the stream, whereas the scaling $\propto t^{-2.4}$ is the steeper scaling predicted for material that is being affected by the presence of the clumps \citep{coughlin19}. }
   \label{fig:fallback}
\end{figure}

Figure \ref{fig:fallback} illustrates the fallback rate $\dot{M}$ onto the black hole from the tidally ejected tail in units of solar masses per second as a function of time in seconds. To measure this quantity, we reran the same simulation that yields the disc-tail structure in Figure \ref{fig:disruption} but artificially increased the accretion radius of the black hole to $20 \,GM_{\bullet}/c^2 \simeq 100$ km at a time of $8.75\times 10^{-3}$\,s. This curve therefore represents the rate at which material from the tail returns to pericenter (or impacts the disc; see discussion below). We checked the debris distribution between this simulation and the full simulation at late times and find no change in the stream structure or evolution. For times $t \lesssim 0.1$ s, the curve in Figure \ref{fig:fallback} shows a smooth decline that is well-approximated by the blue, dashed curve $\dot{M} \propto t^{-5/3}$, which is the scaling expected from the return of material from the marginally bound radius within the tail (e.g., \citealt{chevalier89}). However, for times $t \gtrsim 0.1$ s, the fallback rate starts to exhibit large fluctuations as discrete clumps of material return to the black hole. Moreover, if we approximate the region of the tail in between clumps as dominated by the gravitational field of the black hole and the trailing clump, then using Equation (15) of \citet{coughlin19} and the fact that the ratio of the clump mass to the black hole mass is $\mu \simeq 0.01$ predicts that the return of this material should scale as $\dot{M} \propto t^{-2.4}$. This scaling is shown by the orange, dashed curve in this figure, and provides a good approximation to the average decline (i.e., in between the accretion of clumps) exhibited by the fallback curve.

\begin{figure}
   \centering
   \includegraphics[width=0.475\textwidth]{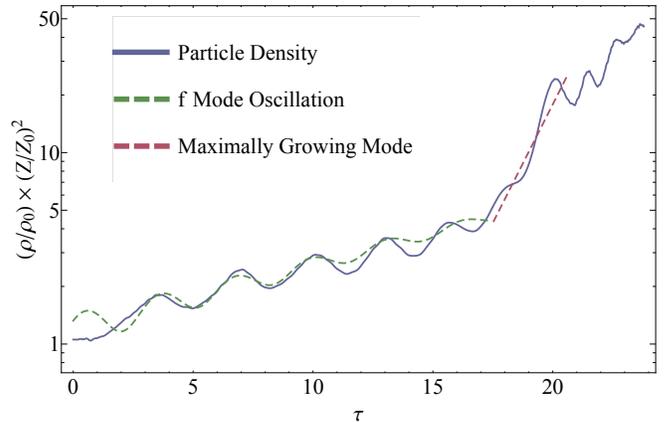} 
   \caption{The blue curve shows, on a log-linear scale, the density of an SPH particle that resides in a knot by the end of the simulation, normalized by its initial density $\rho_0$ and multiplied by $(Z/Z_0)^2$, where $Z$ is the position of the marginally bound orbit; $Z_0$ and $\rho_0$ are both measured at a time of $2.5\times 10^{-3}$ s post-disruption (see the top-left panel of Figure \ref{fig:disruption}) and $\tau$ is a time-like variable (see Equation \ref{tauofZ}). In the absence of an instability, this curve would only show stable oscillations, with an oscillatory frequency given approximately by the $f$-mode of a $\gamma = 2$ polytrope, which is shown by the green, dashed curve. There is a slight growth, however, at early times that is due to the aggregate of weakly growing, very long wavelength perturbations that are all unstable; the slope of the green, dashed curve is chosen to match this slow growth to highlight the agreement between the prediction of the $f$-mode oscillatory frequency and the numerical results. The red, dashed line shows the growth rate predicted for the maximally growing mode, the amplitude of which -- for this simulation -- likely grows primarily out of nonlinear couplings of the more weakly growing, long wavelength perturbations. }
   \label{fig:rhoi_of_tau}
\end{figure}

The blue curve in Figure \ref{fig:rhoi_of_tau} shows the product $\propto \rho \times Z^2$, where $\rho$ is the density of an SPH particle that, by the time appropriate to the bottom panel of Figure \ref{fig:disruption}, belongs to a clump that forms out of the instability and $Z$ is the location of the marginally bound radius within the tail (we have examined particles in several clumps and this behavior is typical of each clump). The quantities $\rho_0$ and $Z_0$, being the initial density and position of the marginally bound orbit, are measured at a time of $t = 2.5\times 10^{-3}$ s (see the top-left panel of Figure \ref{fig:disruption}). The density is plotted as a function of the time-like variable $\tau$, which scales with the center of mass position as in Equation \eqref{tauofZ}. We find for this specific simulation that

\begin{equation}
    \frac{\sqrt{G\Lambda_0}}{H_0}\frac{Z_0^{3/2}}{\sqrt{2GM_{\bullet}}}\frac{1}{\frac{3}{2}-\frac{1}{\gamma-1}} \simeq 7.0, \label{norm}
\end{equation}
which normalizes the definition of $\tau$. As noted at the end of Section \ref{sec:analytics}, the fact that this number is of the order unity is expected based on the fact that the star was successfully disrupted. 

\begin{figure*}[htbp] 
   \centering
   \includegraphics[width=0.52\textwidth]{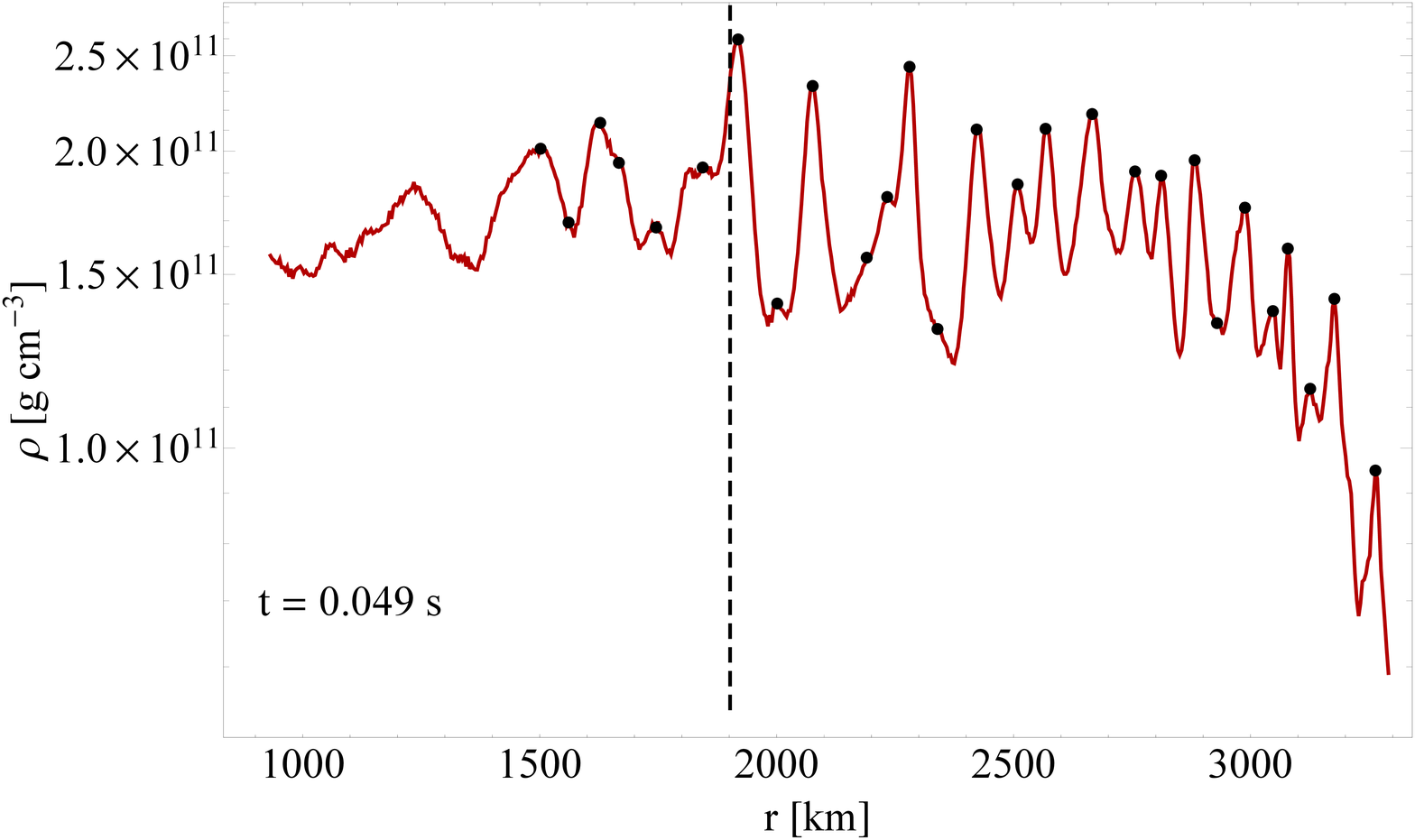} 
   \includegraphics[width=0.465\textwidth]{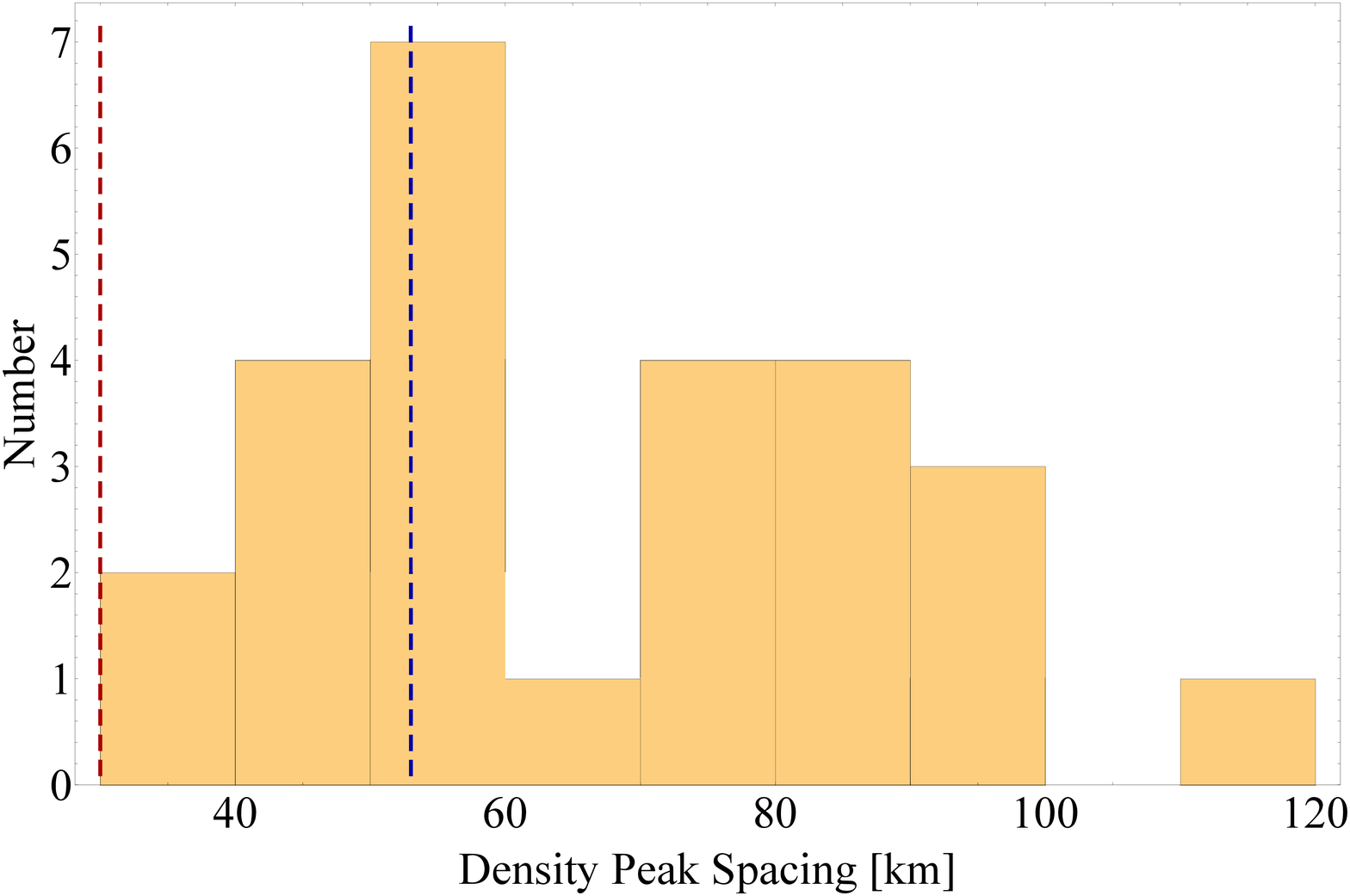}
   \caption{Left: The radial density profile of the stream, averaged over the stream width, at a time of $\sim 0.05$ s post-disruption. Shortly after this time the nonlinear phase of the instability sets in, and the stream fragments into a collection of knots; this time corresponds roughly to $\tau \simeq 17$ in Figure \ref{fig:rhoi_of_tau}, and is therefore predicted to be near the onset of the runaway growth of the most unstable mode. The points coincide with local maxima in the density profile. The vertical, dashed line shows the location of the marginally bound radius within the stream at this time. Right: A histogram of the spacing (in units of km) between the black points in the left panel of this figure, where bin widths were set to 10 km; this bin width is a compromise between over and undersampling the number of points within each bin. The vertical, blue, dashed line is the prediction for the spacing of the density peaks that arises from the maximally growing mode, being $\sim 53$ km for this specific simulation. The vertical, red, dashed line is the minimum spacing of the density maxima, below which perturbations are predicted to be stable and runaway growth should not occur; this spacing corresponds to $\sim 2\pi R_{\star}/k_{\rm crit}$, where $k_{\rm crit} \simeq 1.75$ is the critical wavenumber (relative to the cylindrical radius of the tail) above which perturbations are stable, and is $\sim 29$ km for this simulation.}
   \label{fig:density}
\end{figure*}

If the tidal tail were hydrodynamically stable, then the product $\rho Z^2$ would exhibit gravito-acoustic oscillations at frequencies appropriate to the fundamental modes of an adiabatic cylinder, but the average value of this product would be unchanged. The dashed, green curve in this figure shows a sinusoidal dependence that has a frequency given by the fundamental mode of a $\gamma = 2$, polytropic cylinder, the frequency of which is $\sigma \simeq 1.7$ \citep{coughlin20}. We see that this mode does, indeed, characterize the oscillatory nature of the density perturbations, and we emphasize that the solution does appear to be periodic in $\tau$, implying that the declining sound speed of the stretching stream induces a periodicity in time $t$ that varies as $\sim \sin(\sigma t^{1/3})$. However, there is also a slowly increasing trend exhibited alongside the oscillatory variation (from $\tau =0$ to $\tau \simeq 17$), which scales approximately as $\propto e^{0.07\tau}$ (note that Figure \ref{fig:density} is on a log-linear scale). This growth rate, however, is much smaller than the one corresponding to the maximally-growing mode, being $\propto e^{0.57\tau}$. 

This very slow increase in the product $\rho Z^2$, and the corresponding instability, arises from the fact that the stream possesses an initial perturbation from the polytropic density profile of the star. Specifically, instead of being exactly flat, at early times the density profile of the tidal tail possesses a long-wavelength perturbation from the stretching of the centrally-peaked density profile of the spherical polytrope (i.e., we can approximate the stream density as the spherical density profile of the star stretched in one dimension, which has a maximum near the stellar center of mass).  The majority of the power of the Fourier decomposition of this perturbation is contained at small Fourier wavenumber $k$ (wavelengths much greater than the radius of the tail $H$), all of which are unstable but grow at very slow rates (see Figure 4 of \citealt{coughlin20}). As the stream stretches, all of these modes grow and give rise to the slow increase of the product $\rho Z^2$. 

In addition to growing linearly, however, nonlinear couplings between the modes also transfer power to higher $k$ (smaller wavelengths). Since there is a wavenumber with the fastest growth rate, over time this mode preferentially ``steals'' power from the longer wavelength modes and emerges at later times. We see this behavior at a time of $\tau \simeq 17$, where the growth rate of the instability steepens and is well approximated by the prediction for the fastest-growing mode (the red, dashed line in this figure). 

To support the notion that we are seeing the emergence of the maximally growing mode, the left panel of Figure \ref{fig:density} shows the density profile as a function of radius, binned into $\sim 1000$ radial bins and averaged over the small solid angle of the stream, at $\tau \simeq 17$ (or a time post-disruption of $t = 0.049$ s), which is at the onset of the appearance of the maximal growth rate in Figure \ref{fig:rhoi_of_tau}. Each point corresponds to a local maximum in the density of the stream, and the vertical, dashed line in this figure shows the location of the marginally bound radius at this time (i.e., the fluid element at this location has a Keplerian specific energy of zero). The right panel of this figure shows a histogram of the distance (in km) between successive clumps binned into widths of 10 km. One prediction from the linear perturbation theory \citep{coughlin20} is that the maximally growing mode for a $\gamma = 2$ polytrope occurs at a wavenumber of $k_{\rm max} \simeq 0.97$, and that the spacing between maxima should preferentially be given by $\simeq 2\pi H_0/k_{\rm max} \simeq 53$ km for this simulation; this spacing is shown by the vertical, dashed, blue line, which agrees well with the peak in the histogram. A second prediction is that there is a larger wavenumber (smaller wavelength) $k_{\rm crit} \simeq 1.75$ below which perturbations are stable, and hence density maxima within the stream separated by a length less than $\simeq 2\pi H_0/k_{\rm crit} \simeq 29$ km should be suppressed; this spacing is shown by the vertical, dashed, red line, and does reproduce the observed cutoff in the spacing distribution of the local maxima. 

To summarize and compare the numerical results here with the analytic calculations of Section \ref{sec:analytics}, the simulations here yield a fragmentation time of around $t_{\rm frag} \simeq 50$ ms, while we estimated this time to be $\sim 70$ ms; a preferred spacing between clumps of $\sim 53$ km, compared to the prediction of $\sim 30$ km; and a total number of $\sim 40$ knots that condense out of the stream, which is between the upper and lower limits we anticipated (but much closer to the lower limit of 20, as expected from the fact that a substantial amount of material has accreted by the time fragmentation occurs). The fallback rate in Figure \ref{fig:fallback} also shows that variability begins around a time of 0.1 s, while the predicted time was closer to $\sim 1$ s. The spacing between the spikes in Figure \ref{fig:fallback} is also on the order of $\sim 10 - 100$ ms, which is consistent with the prediction, and there does appear to be an increase in the temporal separation between spikes in the fallback rate as time advances.

\section{Observational Implications}
\label{sec:observational}
The preceding sections demonstrate that the tail of ejecta shed from the destruction of a neutron star during a compact object merger can fragment under its own self-gravity into self-bound knots, and that this fragmentation occurs on a timescale of tens of milliseconds (e.g., Equations \ref{tfrag0} and \ref{tfrag1}). The number of knots formed is generally on the order of tens owing to the short return time of the most bound debris relative to the fragmentation time (e.g., Figure \ref{fig:disruption} and Equations \ref{Nupper} and \ref{Nlower}); the bound knots return to the disrupting object on a timescale of hundreds of milliseconds, while the unbound knots escape from the system on hyperbolic trajectories and at $\sim 10\%$ the speed of light. Here we briefly discuss the observational implications of these findings. 

\subsection{Prompt Emission}
With our simulations that adopted a post-Newtonian prescription for the gravitational potential of the disrupting black hole, the pericenter distance of the neutron star implied that the angular momentum of the returning material was sufficient to form an accretion flow outside of the innermost stable circular orbit (ISCO) of the remnant. As such, the material retained enough angular momentum to circularize and form a large-scale disc that spread viscously over time (Figure \ref{fig:disruption}). As knots returned to pericenter, they impacted the disc and generated shocks, and in our simulation the heat from these shocks was instantaneously radiated. More realistically, however, a large fraction of the heat will be trapped in the accretion flow, which will inflate the disc (though the material may still radiate a significant amount of energy in the form of neutrinos; e.g., \citealt{lee09}) and significantly lengthen the amount of time over which the radiation from such shocks is emitted from the $\tau \simeq 1$ surface; the accretion rate is also extremely super-Eddington\footnote{Adopting a radiative efficiency of $\eta = 0.1$, the Eddington accretion rate of a $5M_{\odot}$ black hole with an electron scattering opacity of $\kappa_{\rm es} = 0.34$ cm$^{2}$ g$^{-1}$ is $\dot{M}_{\rm Edd} \simeq 4\times10^{-15} M_{\odot}$ s$^{-1}$, compared to the values of $\sim 1M_{\odot}$ s$^{-1}$ we obtain for the fallback rate at early times from the simulation; see Figure \ref{fig:fallback}.} and the optical depth is very large to radiation\footnote{The outer disc radius of the simulation in Figure \ref{fig:disruption} is $R \sim 100$ km, and hence a characteristic optical depth is $\tau \simeq \rho\kappa_{\rm es}R \simeq M_{\star}\kappa_{\rm es}/R^2 \simeq 6\times 10^{18}$; this estimate -- which assumes a spherical distribution of material -- is clearly an over (under) estimate outside (inside) of the disc midplane, but is nonetheless a measure of the extreme optical depths encountered in these compact object mergers.} and even to neutrinos in the inner regions (e.g.,\citealt{woosley93, popham99, lee09, siegel18}). Consistent with previous works, we thus conclude that incorporating the effects of the radiation field and neutrinos would likely result in the formation of an optically thick, relativistic outflow, the likes of which can power the prompt emission of a short GRB (e.g., \citealt{goodman86,paczynski86,krolik91}) but would likely hinder the direct detection of the radiation generated by knots impacting the accretion flow.

On the other hand, if the jetted activity thought to be responsible for the gamma-ray production is linked to and directly proportional to the accretion rate, then -- if the viscosity present in the accretion flow is large enough -- the changes in the fallback rate as knots return to pericenter will induce a comparable variability in the accretion rate and central engine power. If the viscosity is large then this variability can occur on timescales as short as the dynamical time within the disc, on the order of $\sim$ tens of ms, consistent with observations of the prompt $\gamma$-ray emission and early X-ray emission (e.g., \citealt{klebesadel73, gehrels06, margutti10, margutti11}). As noted in Section \ref{sec:disc}, we did not employ any explicit physical disc viscosity in our simulation, and correspondingly the rate of viscous accretion is artificially low. Specifically, the numerical viscosity in our simulated disc is much smaller than the viscosity expected from, e.g., the magnetorotational instability (MRI) in fully ionized discs, which observations indicate has a \citet{shakura73} $\alpha \sim 0.3$ \citep{king07,martin19}. 

If such hydromagnetic turbulence is active in these discs, it is also possible that the accretion timescale of the material within the disc could be shorter than the fallback time of the most bound clump or the time in between the return of successive knots. In this scenario, much of the gas will have been depleted from the vicinity of the returning knots, and -- instead of crashing into and merging with the disc -- each knot will be disrupted\footnote{Assuming that the density of the collapsed objects is of the order of or less than the density of the original neutron star; this will always be the case if the gas retains the same polytropic equation of state as the original star, such that the entropy and adiabatic index remain unaltered throughout the disruption.} upon returning to pericenter. These secondary disruption episodes will fuel further accretion onto the compact object, and may ``restart'' the central engine in quasi-periodic bursts (cf.~\citealt{king05}).

Moreover, if the black hole is rotating at an angle that is inclined with respect to the initial orbital plane of the binary, the disc that forms will be inclined with respect to the returning debris (see, e.g., \citealt{stone12,franchini16,ivanov18} for discussion of how this applies to stellar disruption by supermassive black holes). In this case, even if a large-scale disc is present at the time that knots return, the knots may pass through the disc at an oblique angle and at a distance much larger than pericenter; upon reaching pericentre they are disrupted, forming secondary accretion flows that interact with one another. Such debris orbits may interact at different orbital phases, driven by nodal and apsidal precession, leading to shocks and subsequent accretion \citep[cf.][]{nixon12}. Over time the reservoir of gas that builds may, after accounting for the effects of radiation pressure, resemble more of a quasi-spherical envelope that enshrouds the black hole, with bursts of accretion driven by the interaction of the precessing discs that form. We note that if the spin of the accretor is primarily determined by the angular momentum of the recently merged binary, as expected for nearly equal mass mergers, then it is unlikely that the ejected streams will be strongly misaligned with the accretor's spin, but nonetheless a significant spin-debris misalignment may be possible in some cases (especially in black hole-neutron star mergers, where the mass ratio can deviate substantially from unity).

\subsection{Extended Emission}
A number of short gamma-ray bursts also display ``extended emission,'' which is softer $\gamma$-ray/X-ray emission that is not generated by the interaction of ejecta with surrounding material, continues for hundreds of seconds following the prompt burst, and can contain as much as or more energy than the prompt spike of emission (e.g., \citealt{lazzati01, della06, gehrels06, nakar07, perley09, norris11, kisaka17, burns18}). As discussed by other authors (e.g., \citealt{faber06, lee07, metzger10, desai19}), the late-time fallback of material to the black hole (and the continued accretion thereof) is one promising means of producing this emission, and our results here serve to further substantiate this origin.

Additionally, this late-time emission is often highly variable, and on timescales much shorter than the $\sim 100$ second duration of the extended emission itself, which suggests that the emission arises from near the compact object. If the extended emission is indeed fueled by the fallback of weakly bound material from the tidal tail, this variability can be explained by the gravitational instability of the stream identified here and is driven by the return of knots to the compact object. The mechanism responsible for communicating the fallback rate to the black hole, and ultimately how this increased energy and (presumably) magnetic flux at the event horizon translates into an increase in the jet power, dictates the relative fluence of energy between the prompt and extended emission. Depending on how this mechanism operates, this model for the powering of the extended emission may be consistent with scenarios in which the fluence of the extended emission exceeds that in the prompt emission (as identified in, e.g.,  \citealt{gehrels06,perley09}). For example, while the raw fallback rate in Figure \ref{fig:fallback} contains a much larger mass flux at earlier times, which naively translates to a larger accretion luminosity, the outward transport of angular momentum in the disc implies that a large fraction of this matter could be contained in a reservoir that accretes at a later time. In this case, the extended emission would contain an amount of mass (and would liberate an amount of accretion energy) that could conceivably exceed that contained in the prompt emission. 

\subsection{R-process \& Kilonova}
As noted above, the knots that form out of the gravitational instability are distributed over the bound and unbound segment of the tail. When the mass ratio of the inspiraling objects is fairly dissimilar from unity, as is the case for a black hole-neutron star merger considered here, the tidally ejected tail contributes substantially to the r-process production of heavy elements and the corresponding kilonova afterglow (e.g., \citealt{lattimer74, lattimer76, lattimer77, meyer89, rosswog00, metzger10, metzger12, barnes13, barnes16, radice18, tsujimoto20}). Indeed, if the angular momentum of the bound material places the disc within the ISCO of the black hole, the unbound tail provides the \emph{only} source of r-process enrichment. If the unbound tail rapidly (on the order of ms; see Section \ref{sec:analytics}) fragments into a number of distinct knots separated by tenuous material, one would expect qualitative differences in the appearance of the kilonova (compared, e.g., to a spherically symmetric outflow) that occurs days after the disruption, owing to variations in the optical depth along the filament and the reduction in the total emitting surface area. The mixing of the r-process-enriched gas would also be less efficient because of the much smaller effective volume maintained by the ejecta.

\subsection{Afterglow}
Alongside the prompt $\gamma$-ray and early X-ray emission, the interaction between the relativistic ejecta and the circumburst medium should generate emission at longer wavelengths in the form of an X-ray, optical and radio afterglow (e.g., \citealt{rees92, meszaros97, sari98, nakar11, margalit20}; though in some short GRBs the relatively ``clean'' environment of a short GRB, being in the outskirts of a galaxy following a natal kick, correspondingly reduces the brightness of this component of the emission; e.g., \citealt{narayan92, nysewander09}). To the extent that the rate of return of material to the compact object influences the formation and energetics of the relativistic outflow, we would qualitatively expect the sudden enhancement in the accretion rate through the return of a discrete knot to imprint itself on the afterglow, possibly in a way that would mimic the ``refreshed shock'' scenario (e.g., \citealt{rees98, kumar00}).

In addition to the relativistic outflow formed from the central engine, a radio transient should also be generated from the interaction between the less relativistic, unbound tail and the surrounding medium. The fragmentation of this tail into a number of discrete knots greatly reduces its cross-sectional area, which correspondingly dramatically inhibits the production of this distinct radio transient. As for the case of r-process emission, this would be the only source of radio emission if the angular momentum of the fallback disc is insufficient to allow the material to circularize outside of the ISCO.

\subsection{Gravitational Waves}
Finally, in addition to electromagnetic counterparts, the successive return of knots to pericenter will generate a distinct gravitational-wave signal, and will be characterized by a train of ``chirps'' -- occurring as individual knots return to pericenter -- that accompany potential flares in the electromagnetic signal as the knots are disrupted to form secondary accretion flows or impact the disc. The amplitude of the gravitational-wave signal will clearly be reduced dramatically below that of the initial inspiral owing to the much smaller mass ratio, and will only be detectable by current facilities (i.e., LIGO; \citealt{aasi15}) if the event is very nearby ($\lesssim 10$ Mpc), but may be observed by future generation facilities at more reasonable distances. The detection of this concomitant gravitational wave signal would be among the most convincing pieces of evidence to support the existence of this instability operating in short GRBs.

\section{Summary and Conclusions}
\label{sec:summary}
Our analytical arguments (Section \ref{sec:analytics}) demonstrate that, if the gas comprising the tidal tails formed during the merger of two compact objects is adiabatic and has an effective adiabatic index $\gamma$ that satisfies $\gamma \ge 5/3$, then such tails are unstable and fragment globally -- along the axis of the tail -- under their own self-gravity into knots with radii of the order the width of the tail. Our simulations of the disruption of a $2M_{\odot}$ neutron star, modeled as a $\gamma = 2$ polytrope with a radius of 11 km, by a $5M_{\odot}$ black hole that adopt an adiabatic index of $\gamma = 2$ (Section \ref{sec:numerics}) show agreement with the predictions of the spacing of the knots that condense out of the instability (Figure \ref{fig:density}), the number of knots formed, and the linear growth of the instability itself (Figure \ref{fig:rhoi_of_tau}). The return of these knots to the compact object results in variability in the fallback rate (Figure \ref{fig:fallback}) on timescales commensurate with our predictions, being on the order of tens of millseconds.

As described in more detail in Section \ref{sec:observational}, this timescale over which variability occurs in the fallback rate is roughly consistent with observed variability in the prompt emission of short GRBs. Thus, depending on how rapidly these changes in the fallback rate can be communicated to the accreting object and translated to an accretion rate, the return of individual knots to pericenter can plausibly contribute to the flaring in the lightcurves of short GRBs. The return of these knots to pericenter also continues to later times (i.e., much later than the $\sim 1-2$ second duration of the prompt emission), and can generate variability in the ``extended emission'' observed in some sources. Owing to its substantially smaller cross-sectional area, the fragmentation of the unbound tail into discrete knots will also significantly reduce the intensity of the radio transient that forms as the tail slams into the circumburst medium. Finally, the return of discrete knots to pericenter should also create a gravitational-wave signal, consisting of a train of ``chirps,'' though it will likely be very difficult to detect owing to the relatively small amount of mass contained in individual knots. 

Here we focused primarily on the case of a neutron star disrupted by a black hole, as these unequal-mass-ratio encounters result in a significant amount of dynamical ejecta; in our simulation we found that $\sim 10\%$ of the initial star comprises the ejecta, which agrees well with other, more detailed simulations (e.g., \citealt{kyutoku15}). On the other hand, neutron star-neutron star mergers -- for which the mass ratio is generally much closer to unity -- eject less mass ($\sim 0.01 - 0.1 M_{\odot}$) prior to the coalescence of the objects (e.g., \citealt{shibata19}). As we noted in Section \ref{sec:analytics}, the timescale for the instability to develop is directly related to the line mass of the tidal tail, with larger (smaller) amounts of mass yielding shorter (longer) instability timescales. We therefore expect that for short GRB progenitors with mass ratios closer to unity, such as neutron star-neutron star mergers, the instability identified here will take longer to develop and the mass contained in the knots will be smaller, which in general will make its presence more difficult to detect.

\acknowledgements
We thank Jim Pringle and Adam Burrows for useful discussions, and Gavin Lamb for providing comments on an early draft. We thank the anonymous referee for useful and constructive comments and suggestions. We thank William Lee for giving us useful comments and for drawing our attention to the work of \citet{lee00}. ERC acknowledges support from NASA through the Hubble Fellowship, grant No.~HST-HF2-51433.001-A awarded by the Space Telescope Science Institute, which is operated by the Association of Universities for Research in Astronomy, Incorporated, under NASA contract NAS5-26555. CJN is supported by the Science and Technology Facilities Council (grant number ST/M005917/1), and also acknowledges funding from the European Union’s Horizon 2020 research and innovation program under the Marie Sk\l{}odowska-Curie grant agreement No 823823 (Dustbusters RISE project).
J.B. is supported by the National Aeronautics and Space Administration (NASA) through the Einstein Fellowship Program,  grant number PF7-180162.
BDM gratefully acknowledges support from the Simons Foundation (grant \#606260) and the NASA Astrophysics Theory Program (NNX17AK43G).  This work was performed using the DiRAC Data Intensive service at Leicester, operated by the University of Leicester IT Services, which forms part of the STFC DiRAC HPC Facility (www.dirac.ac.uk). The equipment was funded by BEIS capital funding via STFC capital grants ST/K000373/1 and ST/R002363/1 and STFC DiRAC Operations grant ST/R001014/1. DiRAC is part of the National e-Infrastructure. We used {\sc splash} \citep{price07} for some of the figures. 

\software{{\sc phantom} \citep{price18}; {\sc splash} \citep{price07}}

\bibliographystyle{aasjournal}
\bibliography{refs}

\appendix
\section{Caveats of the Numerical Approach}
\label{sec:caveats}
Here we discuss some of the caveats and assumptions of our numerical models, and we speculate as to the impact of relaxing some of these assumptions.

\subsection{Disc physics}
\label{sec:disc}
The neutron star material that is deeper in the gravitational potential well of the black hole (at the time it crosses the tidal radius) promptly forms an accretion disc as it precesses relativistically and intersects itself. Over time, material from the returning tail feeds this disc, and it grows in radial extent as the angular momentum budget increases. By the end of the simulation ($\sim 2$ seconds), the disc is extended in radius out to $\sim 100$ km, and knots returning to the black hole impinge upon this disc as reflected in the fallback rate (Figure \ref{fig:fallback}).

We emphasize, however, that our model of the disc is likely not physically appropriate to the extreme conditions under which the disc forms and evolves. For one, our equation of state assumes that any heat generated from the production of shocks -- which predominantly mediate the disc formation as general relativistic apsidal precession causes the material to self-intersect (e.g., \citealt{rosswog02}) -- can be efficiently radiated from the system, whereas the optical depths are so large that this energy should be trapped and heat the flow. The temperatures and densities are also so high that neutrino cooling and even nuclear burning are not negligible, both of which will modify the thermodynamics and feed back onto the fluid dynamics of the flow (e.g., \citealt{fernandez16}). Indeed, the densities are so high at early times that neutrinos can couple efficiently to the gas, generating a neutrino-driven wind (e.g., \citealt{fernandez13, siegel18}). Our disc, on the other hand, retains a polytropic equation of state, while these other effects would cause the disc to puff up in the vertical direction and have an associated outflow.

Our simulations also do not employ magnetic fields, which likely lead to additional dissipation through the existence of the magnetorotational instability \citep{balbus91}. This instability may manifest itself as an increase in the effective viscosity coefficient $\alpha$ that controls the rate at which angular momentum is transported within the disc \citep{shakura73}. The viscosity present in our simulations, by contrast, is only at the numerical level, which is small for the large number of particles ($N_{\rm p}\gtrsim 10^7$) we used. As a consequence, the accretion rate of the black hole is unrealistically small in our simulations, and the fallback rate in Figure \ref{fig:fallback} is likely a better approximation of the mass flux at the event horizon (modulo a suitable time lag) in a realistic disc that has $\alpha \sim 1$.

Our simulations also do not account for radiative processes; while the densities and temperatures are initially so high that neutrino cooling and heating are the dominant form of the transport of accretion energy (e.g., \citealt{popham99}), eventually the temperature and density in the inner disc regions will drop to the point where neutrino production is inefficient. In this case, the dominant transport mechanism will be the advection of radiation throughout the disc, but the accretion luminosity will be so well in excess of the Eddington limit of the black hole (even for extremely small radiative efficiencies) that it is difficult to see how outflows will be avoided (e.g., \citealt{blandford99}). The radiation will also be trapped within the flow \citep{begelman78}, leading to -- in addition to outflows -- a much more vertically extended disc structure than we find here.

\subsection{General relativity}
\label{sec:gr}
Our simulations approximate the gravitational field of the black hole with a Paczynski-Wiita potential; in particular, we employ $-GM_{\bullet}/(r-R_{\rm S})$ for the potential of the black hole, where $R_{\rm S}$ is the Schwarzschild radius of the black hole. This approach is clearly much more simplistic than solving the Einstein equations, which is the methodology employed by very recent simulations of the mergers of compact objects (e.g., \citealt{bauswein19,foucart19}).

Solving the Einstein equations for the dynamical evolution of the spacetime is necessary for accurately modeling the disc physics in the regions in the immediate vicinity of the black hole event horizon. It is also necessary for understanding the initial, tidal deformation of the neutron star, as for our setup -- a $5M_{\odot}$ black hole and a $2M_{\odot}$ neutron star with a radius of 11 km -- the fiducial tidal radius is comparable to its gravitational radius. Indeed, the pericenter distance of the star that leads to the disruption in Figure \ref{fig:disruption} was $3 R_{\rm G}$, which is likely to be stretching the accuracy of the Paczynski-Wiita potential (see, e.g., Figure 4 of \citealt{tejeda13}).

Including general relativity would likely strengthen the tidal interaction that results in the disruption of the star, as it is generally the case that pseudo-Newtonian and effective potentials underpredict general relativistic quantities (e.g., the relativistic advance of periapsis, which can be thought of as an additional term that enhances geodesic deviation, is generically underpredicted by effective potentials; e.g., \citealt{tejeda13}). Consequently, it is likely that a general relativistic treatment, or including a potential that maintains higher-order corrections in $r / R_{\rm S}$, would not require as close a pericenter distance to completely disrupt the star (see Section \ref{sec:orbital}); for the disruption of Solar-like stars by supermassive black holes, a partial disruption with a Newtonian treatment of the gravitational field of the supermassive black hole may become a full disruption when one uses general relativity \citep{gafton15,gafton19, stone19}. 

Properly accounting for relativity would clearly enhance the physicality of the initial disruption and the disc physics. However, this would also add significantly to the computational cost of the simulation, and would not be essential for understanding the fragmentation of the tidal tails -- the main focus of this paper -- which occurs at hundreds to thousands of gravitational radii of the black hole.

\subsection{Microphysics and thermal physics of the tidal tails}
\label{sec:thermal}
The evolution of the gas in our simulation is adiabatic with a polytropic index of $\gamma = 2$. As discussed in Section \ref{sec:numerics}, this choice was motivated by the theoretical work on the nuclear equation of state, which suggests that the effective adiabatic index of the neutron star material is $\gamma \sim 2-3$, and that more accurate modeling seems to indicate that piecewise-polytropic functions can reproduce more complicated equations of state \citep{read09}. This approach is also identical to what appears to have been done in many previous investigations of this problem (e.g., \citealt{lai94, lee99, rosswog99, lee01, faber06, lee07, ruiz20}), and we wanted to test the prediction that these tails are subject to gravitational fragmentation if this thermodynamic simplification is made. 

More realistically, however, the decline of the density of the tidal tail prior to the nonlinear phase of the instability (see Figure \ref{fig:rhoi_of_tau}) implies that the extremely high densities characteristic of the nuclear equation of state $\rho \sim 10^{14}$ g cm$^{-3}$ -- and correspondingly the high value of $\gamma$ -- will no longer be maintained. As the density declines, it becomes energetically favorable for neutrons and protons to drip out from the denser regions of the stretching material, and $\beta$ decay and nuclear fission of extremely neutron-rich nuclei result in the formation of less heavy elements alongside the neutron fluid; this also reduces the overall neutron to proton fraction. As this occurs the nucleons and electrons start to behave more classically and possess an adiabatic index that is better approximated between $4/3$ and $5/3$. The equation of state described by \citet{lattimer91}, \citet{shen98}, and those discussed in \citet{shapiro83} exhibit this general behavior (see, e.g., Figure 5 of \citealt{rosswog02}), as do more recent equations of state (e.g., \citealt{shen11, hempel12, steiner13, banik14}). 

These equations of state and the corresponding adiabatic exponent describe the variation of the nucleon-electron pressure with respect to the density. As the density declines and the nuclei decay into lighter elements, the neutrinos become optically thin to  processes such as neutrino-neutrino, neutrino-electron, and neutrino-necleon scattering (and a host of others; see, e.g., \citealt{bruenn85,burrows06}). When the neutrinos are no longer trapped within the flow, a significant amount of energy is lost from the expanding debris tail (e.g., \citealt{sekiguchi16, vincent20}), which correspondingly reduces the pressure below the isentropic value. Since the width of the tail is governed by the balance between pressure and self-gravity, the width of the tail is reduced. However, the sound speed still declines predominantly from the stretching of the tail in the direction of the remnant (see Equation \ref{cs}), which implies that the sound crossing time over the width of the tail is reduced as the tail cools from neutrino emission. This reduction in the sound crossing time implies that the effective $\gamma$ increases as the tail radiates energy in the form of neutrinos.

We conclude that the effective $\gamma$ will soften below the nuclear value as the density declines, but will also stiffen as the material becomes optically thin to neutrinos and energy (and pressure support against self-gravity) is lost. We experimented with changing the equation of state by modifying the polytropic index of the gas (equal to the polytropic index of the initial star). We found -- consistent with the predictions of Section \ref{sec:analytics} and with past investigations \citep{lee07} -- that increasing the adiabatic index resulted in more vigorous fragmentation at earlier times (see also \citealt{coughlin16a}). As we softened the adiabatic index, the fragmentation was less pronounced and occurred later. We did not perform simulations with an adiabatic index $\gamma < 5/3$ to test the prediction that such a configuration is stable to gravitational fragmentation, though the simulations of \citet{coughlin16a}, who studied the structure of debris streams produced from tidal disruption events with a range of $\gamma$, found that fragmentation did not occur for $\gamma < 5/3$. 

Finally, we emphasize that the stability analysis, and the condition that $\gamma = 5/3$ is the critical adiabatic index that separates stable from unstable streams, results from an expansion of the fluid equations about the marginally bound radius. In particular, the spreading of the material in this region causes the sound speed to decline as $\sim 1/Z$, whereas the width of the tail expands subsonically until $\gamma = 5/3$, which allows the stream to remain causally connected in the transverse direction. However, there are higher-order terms (in the quantity $\Delta z/Z$) that enter the fluid equations and modify the stability criterion at late times once we start to move away from the marginally bound radius. For the unbound tail, these additional terms \emph{further destabilize} the stream to self-gravity, as the fluid elements asymptotically approach constant velocities at late times. In this constant-velocity limit, the sound speed declines only as $Z^{-1/2}$, which implies that the unbound segment of the tail is susceptible to gravitational fragmentation until $H \propto Z$; this scaling occurs when $\gamma = 4/3$ (see also Section 5.2 of \citealt{coughlin16}), and demonstrates that the unbound segment of the tail is asymptotically unstable to fragmentation even in the presence of softer equations of state.

\subsection{Resolution}
\label{sec:resolution}
The time at which the linear, gravitational instability reaches the nonlinear regime -- and leads to the formation of knots within the tidal tail -- depends on the magnitude of the initial perturbation that seeds the instability. In our simulation, the dominant source that contributes to the seed density fluctuations is the polytropic density profile of the initial star. In particular, instead of the density along the axis of the tidal tail being completely flat, the geometric center of the stream possesses a slightly increased density relative to the extremities. Consequently, there is an initial density perturbation along the cylinder axis, which gives rise to a collection of simultaneously growing, long-wavelength perturbations. As we argued above, the growth of these modes is likely responsible for the slow increase in the product $\rho Z^2$ in Figure \ref{fig:rhoi_of_tau}, and their nonlinear couplings give rise to additional power at the maximally growing mode that eventually emerges.

In addition to the magnitude of the seed perturbation, the time at which the nonlinear growth is reached also depends on the width of the stream, as it is the dimensionless sound crossing time over this width -- being proportional to the width itself -- that characterizes the oscillatory and growing nature of the perturbations. In the limit of infinite resolution, the width of the tidal tail is determined by the solution to the cylindrical Lane-Emden equation (see Section \ref{sec:analytics}) and is where self-gravity causes the density to equal zero. With a finite number of particles, however, the density cannot ever equal zero, and the stream width is characterized by a location of finite density and pressure. This finite pressure is larger for fewer numbers of particles as the pressure smoothing length is larger, which feeds back on the structure of the tail and correspondingly reduces its width (i.e., the tail becomes effectively pressure confined, which has a smaller equilibrium width than it would if it were in vacuum). Consequently, the growth timescale of the instability as calculated with the numerical method is artificially short, and simulations with fewer particles lead to increasingly shorter timescales in a way that scales linearly with the width of the tail.

\begin{figure*}
    \centering
    \includegraphics[width=0.495\textwidth]{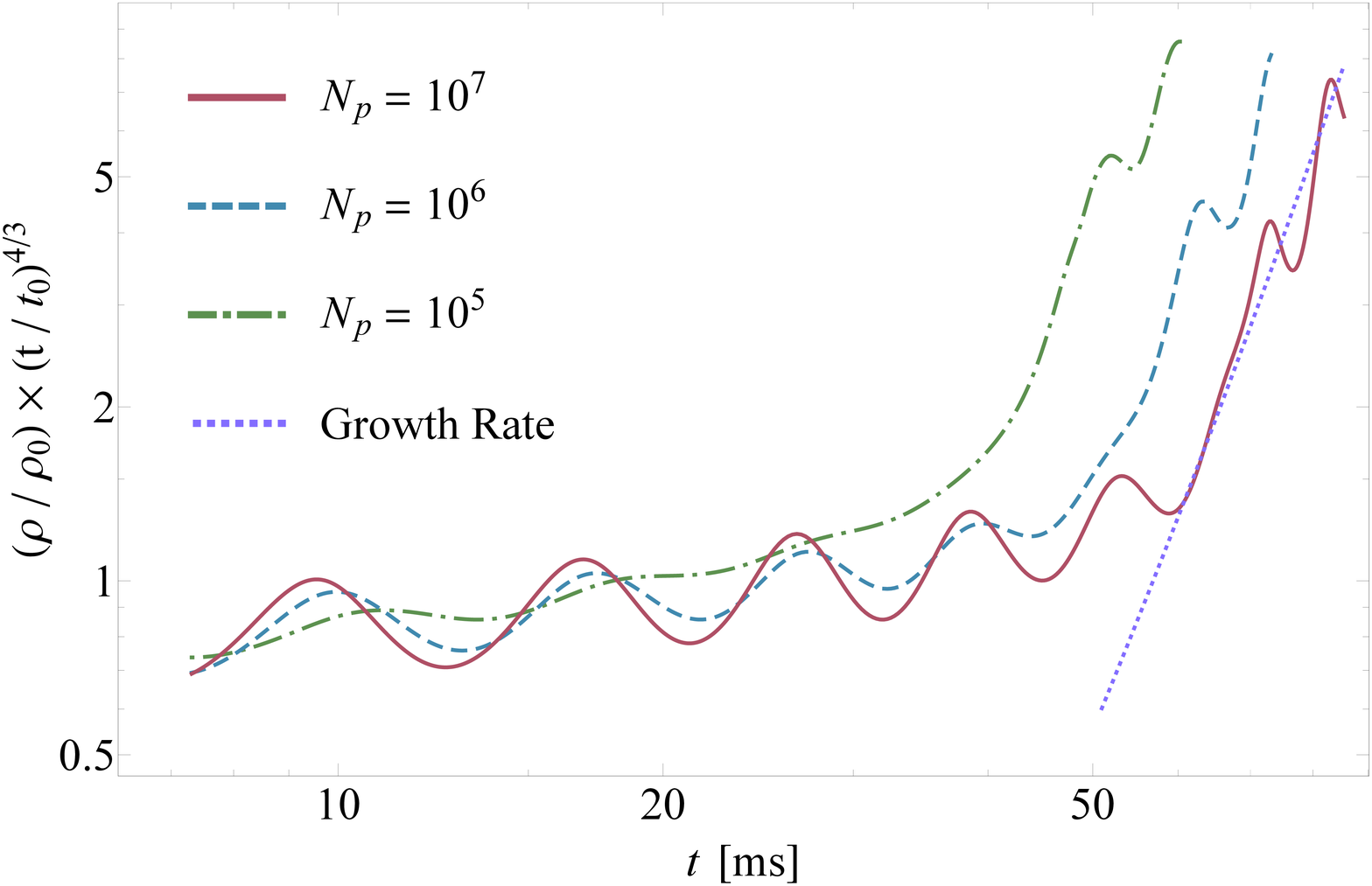}
    \includegraphics[width=0.495\textwidth]{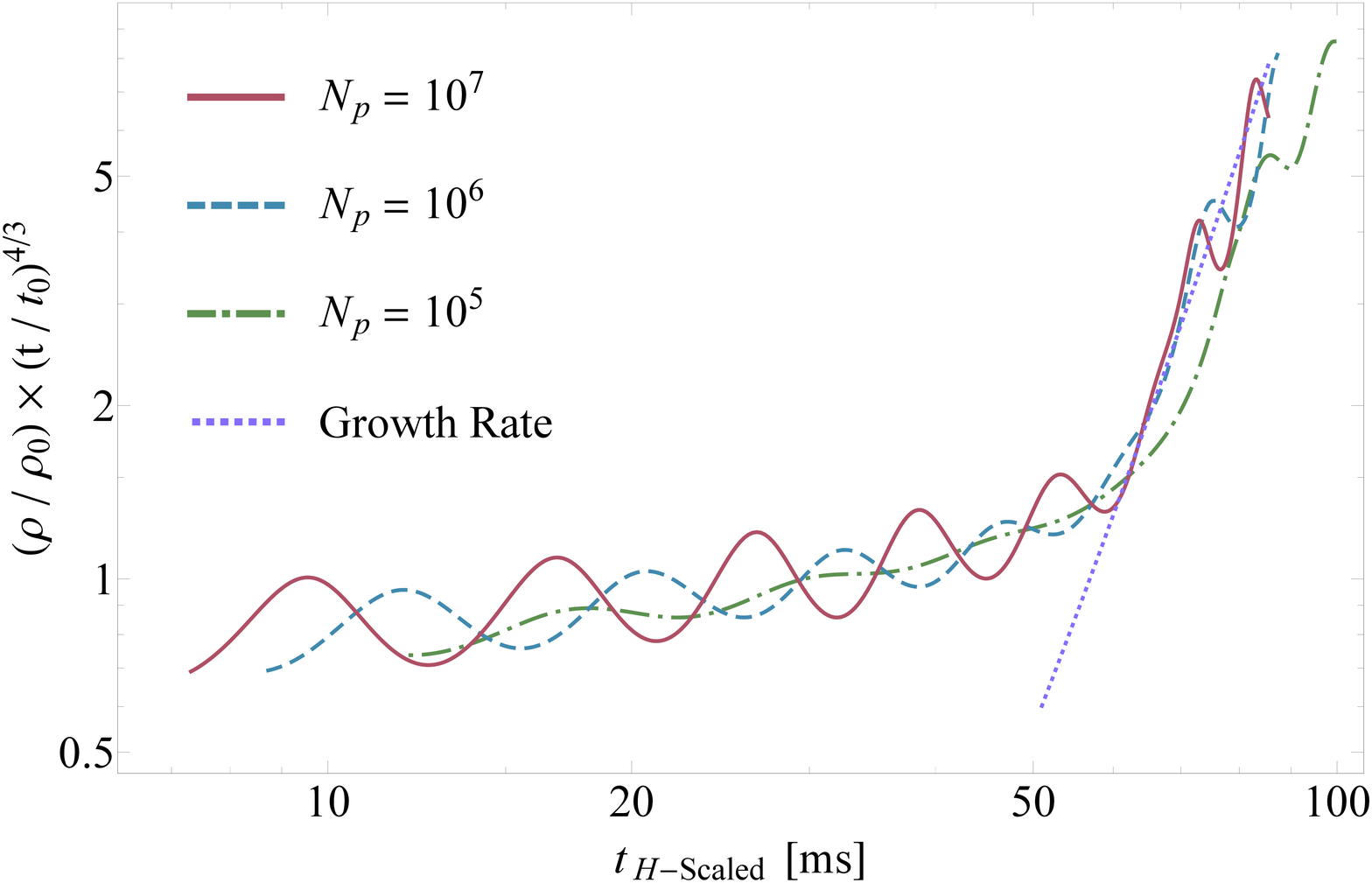}
    \caption{Left: The product $\rho/\rho_0\times(t/t_0)^{4/3}$, where $\rho$ is the average density of the particles that constitute a bound clump that fragments as a result of the instability by the end of the simulation, and $t$ is the time since pericenter passage. Here $t_0 = 2.5$ ms and $\rho_0$ is measured at that time (see the top-left panel of Figure \ref{fig:disruption} for the distribution of the debris at that time). The solid, dashed, and dot-dashed curves are appropriate to the particle numbers $N_{\rm p}$ shown in the legend, and the dotted curve is the growth rate predicted for the maximally growing mode; here the instability shown by the dotted curve grows as $\sim \exp(0.57\times 7.0(t/t_0)^{1/3})$, where $0.57$ is the maximum growth rate predicted from the stability analysis, the factor of 7.0 arises from the ratio of the sound crossing time over the cylindrical radius of the tail to the dynamical time at the location of the marginally bound radius in this simulation (see Equation \ref{tauofZ} and \ref{norm}), and $t_0 \simeq 1.2$ ms is the dynamical time at 2.5 ms post-disruption. This panel shows that the time at which the fastest-growing mode appears changes as a function of resolution. Right: The same as the left panel, but here the time for the $N_{\rm p} = 10^5$ and $N_{\rm p} = 10^6$ particle runs is normalized by the ratio of the stream width appropriate to each simulation to that of the $N_{\rm p} = 10^7$ run. This panel shows that, after accounting for this effect that normalizes the sound crossing time over the width of the stream, the emergence of the most unstable mode appears independently of resolution. This panel demonstrates that, while the resolution affects the time at which the instability manifests itself, the source of the perturbation that seeds the instability is not numerical noise. }
    \label{fig:resolution}
\end{figure*}

To understand how this effect modifies the growth of the instability and the emergence of the most unstable mode, we re-simulated identical disruptions as shown in Figure \ref{fig:disruption} with $N_{\rm p} = 10^5$ and $10^6$. In all three (i.e., including $N_{\rm p} = 10^7$) simulations we located a clump that collapsed out of the stream that was near the marginally bound orbit, and computed the average density of the particles constituting that clump. The left panel of Figure \ref{fig:resolution} shows the evolution of the average density for $N_{\rm p} = 10^5$ (green, dot-dashed), $10^6$ (blue, dashed), and $10^7$ (red, solid) as a function of time in ms (note that this is on a log-log scale and plotted as a function of time, as compared to Figure \ref{fig:rhoi_of_tau}, which is on a log-linear scale and plotted as a function of dimensionless $\tau$). The dotted, light purple line shows the expected growth from the maximally growing mode normalized to the values appropriate to this simulation, i.e., this curve scales as $\sim e^{0.57\times 7.0\times (t/t_0)^{1/3}}$, where $0.57$ is the maximum dimensionless growth rate, the factor of 7.0 comes from the ratio of the dynamical time at the tidal radius to the sound crossing time over the width of the stream at 2.5 ms post-disruption (see Equations \ref{tauofZ} and \ref{norm}), and $t_0 = 2Z_0^{3/2}/(3\sqrt{2GM_{\bullet}}) \simeq 1.1$ ms, where $Z_0 \simeq 160$ km is the location of the marginally bound radius at 2.5 ms post-disruption. We see that, as we increase the resolution of the simulation, the time at which the maximally growing mode appears is extended to later times.

The right panel of this figure illustrates the same three curves, but the time for each simulation is now scaled by the ratio of the stream width of the respective simulation to the width measured from the $N_{\rm p} = 10^7$ run. We find that this ratio is $\sim 0.61$ for $N_{\rm p} = 10^5$ and $\sim 0.83$ for $N_{\rm p} = 10^6$ (i.e., the width of the tail in the $N_{\rm p} = 10^5$ run is roughly 0.61 times the width of the tail in the $N_{\rm p} = 10^7$ run), and hence the time for the $N_{\rm p} = 10^5$ ($10^6$) run is scaled by $1/0.61 \sim 1.65$ ($1/0.83 \sim 1.2$). This figure demonstrates that, by accounting for the sound-crossing time over the width of the stream, the simulations converge in the time taken for the most unstable mode to emerge.

In addition to the physical perturbation that arises from the density profile of the original star, there is a second source that is due to the finite number of particles employed by the numerical method. Specifically, there will always be an inherent level of noise in the density distribution of the tidal tail at the level of the SPH smoothing length, and this noise gives rise to an effective perturbation on that length scale. Initially this scale is much smaller than the width of the stream, and therefore the oscillations induced by these numerical perturbations are stable (i.e., the wavenumber of the perturbation $k$ satisfies $k \gg k_{\rm crit}$, where $k_{\rm crit} \simeq 1.75$; see Figure 4 and Table 2 of \citealt{coughlin20}). However, over timescales much longer than the sound crossing time over the width of the tidal tail, these perturbations are stretched out and their effective $k$ decreases. Therefore, at some time following the disruption of the star, these perturbations will start to ``leak'' into the unstable region of Fourier space, and will non-physically -- and in a way that depends exclusively on the resolution of the simulation -- drive the instability at the fastest-growing mode. If the resolution of the simulation is not high enough, the time taken for the numerical perturbations to leak into the unstable regime will be shorter than the time taken for the long-wavelength modes seeded by the polytropic density profile to nonlinearly couple, and the numerical result will not be converged. \citet{coughlin15} suggested that this noise was ultimately responsible for driving the fragmentation of the debris streams produced from tidal disruption events.

We can estimate the time at which finite resolution will start to artificially drive the instability: from Equation \eqref{vzapp}, the distance between two SPH particles on either side of the marginally bound radius grows approximately as

\begin{equation}
    \Delta z = \Delta z_0\left(\frac{Z}{Z_0}\right)^2.
\end{equation}
If the two particles are originally separated by the SPH smoothing length $h$, then this separation will grow to the radius of the cylinder $H$ -- and will drive growing perturbations -- after the center of mass position reaches

\begin{equation}
    \frac{Z}{Z_0} \simeq \sqrt{\frac{H}{h}} \simeq N_{\rm p}^{1/6},
\end{equation}
where in the last line we used the fact that the interparticle separation is of the order $N_{\rm p}^{1/3}$. Using the fact that $Z \propto t^{2/3}$, if particle noise is seeding the instability, then the time at which the maximally growing mode appears will scale approximately as $N_{\rm p}^{1/4}$.

For our simulations, this scaling of the time for particle noise to influence the fragmentation implies that each increase in the particle number by a factor of 10 should result in a delay of the appearance of the fastest-growing mode by a factor of $10^{1/4} \simeq 1.8$. From Figure \ref{fig:resolution} the time at which the fastest-growing mode appears in the $N_{\rm p} = 10^7$ run is roughly $\sim 60$ ms, and hence the $N_{\rm p} = 10^6$ and $N_{\rm p} = 10^5$ runs should -- if particle noise seeds the instability -- fragment at times of $\sim 60/1.8 \sim 33$ ms and $\sim 60/1.8^2 \sim 19$ ms. Comparing these predictions to Figure \ref{fig:resolution}, it is clear that the fastest-growing mode appears significantly later than it would if the seed perturbations were provided purely by numerical noise, and that this effect is not dominant in contributing to the gravitational fragmentation observed in our simulations.

\subsection{Variation of simulation parameters}
\label{sec:orbital}
In this paper we focused primarily on the results of a simulation in which a $2M_{\odot}$ neutron star, modeled as a $\gamma = 2$ polytrope with a radius of 11 km, was disrupted by a $5M_{\odot}$ black hole. The pericenter distance of the center of mass of the star was equal to $3R_{\rm G}$, where $R_{\rm G}$ is the gravitational radius of the black hole. This simulation reproduced a particularly clean set of initial conditions to study the fragmentation of the tidal tails produced from the disruption: the neutron star was completely disrupted on its initial passage and formed a single, extended tail containing $\sim 10\%$ of the mass of the star.

The mergers that result in short GRBs clearly have a more varied set of initial conditions than this one case study. As such, we performed additional simulations in which we varied the pericenter distance of the stellar center of mass, the mass of the star, the mass of the black hole, the polytropic index of the star and fluid (we tried $\gamma = 1.8$ and 3), the strength of the gravitational field (we replaced the Paczynski-Wiita potential with a Newtonian potential), and -- because the neutron star is likely rotating substantially if the merger happens after the more gradual inspiral of the binary -- the stellar rotation. The most substantial difference generated by varying these parameters was the increased or reduced survivability of the neutron star during its passage. In particular, many of the simulations we performed resulted in a partially disrupted star, which then returned to the black hole at least once to be redisrupted (see also \citealt{rosswog02, rosswog04}). However, in every case we simulated we recovered the formation of at least one tidal tail of debris, which subsequently fragmented under its own self-gravity.

\end{document}